\documentclass[fleqn,usenatbib]{mnras}

\usepackage{newtxtext,newtxmath}
\usepackage{listings}
\usepackage{xcolor}
\usepackage{subcaption}
\usepackage{float}
\usepackage{parskip}
\usepackage{xcolor}
\usepackage{soul}
\usepackage{comment}

\definecolor{codegreen}{rgb}{0,0.6,0}
\definecolor{codegray}{rgb}{0.5,0.5,0.5}
\definecolor{codepurple}{rgb}{0.58,0,0.82}
\definecolor{backcolour}{rgb}{0.95,0.95,0.92}

\lstdefinestyle{mystyle}{
    backgroundcolor=\color{backcolour},   
    commentstyle=\color{codegreen},
    keywordstyle=\color{magenta},
    numberstyle=\tiny\color{codegray},
    stringstyle=\color{codepurple},
    basicstyle=\ttfamily\footnotesize,
    breakatwhitespace=false,         
    breaklines=true,                 
    captionpos=b,                    
    keepspaces=true,                 
    numbers=left,                    
    numbersep=5pt,                  
    showspaces=false,                
    showstringspaces=false,
    showtabs=false,                  
    tabsize=2
}

\usepackage[T1]{fontenc}
\usepackage{fontawesome5}
\usepackage{hyperref}

\setcitestyle{notesep={ }}

\usepackage{graphicx}	% Including figure files
\usepackage{subcaption}	% 
\usepackage{siunitx}   % units
\usepackage{multirow}
\usepackage{amsmath,amstext}
\usepackage{threeparttable}
\usepackage{multicol}
\usepackage{pdflscape} % Advanced maths commands
\usepackage{xcolor} 
\usepackage{siunitx}

\usepackage[nameinlink]{cleveref}
\usepackage{siunitx}
\usepackage{placeins}

\newcommand{\lya}{Ly$\alpha$} 
\newcommand{\snr}{\text{SNR}_{10}}
\newcommand{\github}[1]{%
   \href{#1}{\faGithubSquare}%
}
\DeclareRobustCommand{\VAN}[3]{#2}
\let\VANthebibliography\thebibliography
\def\thebibliography{\DeclareRobustCommand{\VAN}[3]{##3}\VANthebibliography}
\defcitealias{wolfson2023forecasting}{W2023} 
\title[NN Emulator for \lya~autocorrelation]{Neural network emulator to constrain the high-$z$ IGM thermal state from Lyman-$\alpha$ forest flux autocorrelation function}

\author[Z. Jin et al.]{
Zhenyu Jin,$^{1}$\thanks{E-mail: zhenyujin@ucsb.edu}
Molly Wolfson,$^{1}$
Joseph F. Hennawi,$^{1,2}$
and Diego González-Hernández$^{1}$
\\
$^{1}$Department of Physics, University of California, Santa Barbara, CA 93106, USA\\
$^{2}$Leiden Observatory, Leiden University, Niels Bohrweg 2, 2333 CA Leiden, Netherlands\\
}

\date{Accepted XXX. Received YYY; in original form ZZZ}

\pubyear{2024}

\begin{document}
\label{firstpage}
\pagerange{\pageref{firstpage}--\pageref{lastpage}}
\maketitle

\begin{abstract}
We present a neural network emulator to constrain the thermal parameters of the intergalactic medium (IGM) at $5.4 \le z \le 6.0$ using the Lyman-$\alpha$ (\lya) forest flux autocorrelation function. Our auto-differentiable \texttt{JAX}-based framework accelerates the surrogate model generation process using approximately 100 sparsely sampled \texttt{Nyx} hydrodynamical simulations with varying combinations of thermal parameters, i.e., the temperature at mean density $T_0$, the slope of the temperature–density relation $\gamma$, and the mean transmission flux $\langle F \rangle$. We show that this emulator has a typical accuracy of 1.0\% across the specified redshift range. Bayesian inference of the IGM thermal parameters, incorporating emulator uncertainty propagation, is further expedited using \texttt{NumPyro} Hamiltonian Monte Carlo. We compare both the inference results and computational cost of our framework with the traditional nearest-neighbor interpolation approach applied to the same set of mock \lya~flux. By examining the credibility contours of the marginalized posteriors for $T_0, \gamma, \text{and}~\langle F \rangle$ obtained using the emulator, the statistical reliability of measurements is established through inference on $100$ realistic mock data sets of the autocorrelation function. 
%% JFH last sentence here is superfluous. 
% The entire framework is implemented in  and has been made publicly available on GitHub. \href{https://github.com/enigma-igm/igm_emulator.git}{\faGithubSquare}  
\\
\end{abstract}


%%%%%%%%%%%%%%%%%%%%%%%%%%%%%%%%%%%%%%%%%%%%%%%%%%

%%%%%%%%%%%%%%%%% BODY OF PAPER %%%%%%%%%%%%%%%%%%

\section{Introduction}

The epoch of reionization ends the dark ages of the universe and represents one of the most pivotal phases of evolution of the early universe. Photons from the first luminous sources reionized the neutral hydrogen atoms (\ion{H}{I}) in the diffuse intergalactic medium (IGM), driving the creation of first stars, galaxies, and black holes. Understanding the physical process of this transformation remains an open question in cosmological studies. Although the current constraints imply this happened at $z_{\mathrm{reion}} = 6.4 - 9.0$ \citep[$x_{\ion{H}{I}} = 5\% - 95\%$, see][]{planck_2020} where the midpoint is inferred $z_{\text{reion,mid}} = 7.7 \pm 0.7$ from the cosmic microwave background (CMB) observations \citep{planck_2018}, the exact timing, driving sources, and impact on the thermal state of the IGM still remain largely uncertain.

A primary probe of diffuse baryons in the IGM at high $z$ is the \ion{H}{I} Lyman-$\alpha$ (\lya) forest \citep{gunn_peterson_1965, lynds_1971}, which measures the \lya~absorption of intergalactic neutral hydrogen along sightlines to luminous high-$z$ quasars. \lya~transmission measurements suggest
%% JFH suggested --> suggest
the reionization is %was 
%% JFH this shoul dnot be past tense
not complete until $z<6$ \citep{fan_2006, becker_2015, bosman_2018, eilers_2018, yang_2020, bosman_2021_data}. The current observations generally point towards a "late and fast" reionization period \citep{planck_2020},
%% JFH references for this statement?
but their interpretation is obscured by a poor understanding of details of the process. The limited number of direct observational constraints arises from the rapid increase in the \lya~opacity with redshift, which results in the close-to-zero \lya~transmission at $z > 5$ \citep{becker_2015, bosman_2018, eilers_2018}.

To gain further insight from an indirect method, we look at the thermal history of the IGM at $z>5$ \citep{boera_2019, walther_2019, gaikwad_2020}. Photo-heating of neutral atoms during reionization
%% JFH during the reionization --> during reionization 
increases the temperature of the IGM, leading to  
%% JFH the thermal imprints --> thermal imprints
thermal imprints that can shed light on
%% JFH can characterize --> shed light on
the process of reionization and the nature of the ionizing sources \citep{McQuinn_2011, becker_2015, Davies_2018, Kulkarni_2019b}. At small scales, the \lya~forest is sensitive to the thermal state of the IGM due to the following two factors: thermal motions contributing to Doppler broadening of the absorption features and Jeans (pressure) smoothing of the gas altering the underlying baryon distribution while depending on the integrated thermal history of the IGM \citep{Haehnelt_1998, Gnedin_1998,Rorai_Hennawi_2013, Kulkarni_2015, Rorai_Hennawi_2017, onorbe_2017}. 
%%JFH Please add citations to my two papers wtih Alberto Rorai here. 
After the photoionization heating from reionization, the IGM went through a cooling process consisting of Compton cooling due to inverse Compton scattering off CMB photons and %adiabatic cooling due to the expansion of the universe 
%% JFH cooling due to the adiabatic expansion 
cooling due to the adiabatic expansion \citep{McQuinn_2016}. The balance between heating and cooling is anticipated to produce a power-law relationship between temperature and density for low-density gas, expressed as
\begin{equation} 
    T = T_0 \Delta^{\gamma - 1}.
    \label{eq:temperature-density relation}
\end{equation}
Where $\Delta = \rho / \bar{\rho}$ is the overdensity, $\bar{\rho}$ is the mean density of the universe, $T_0$ is the temperature at mean density, and $\gamma$ is the slope of the relationship \citep{hui_1997}. Right after the reionization of \ion{H}{I} ($z \lesssim 6$), $T_0$ is 
%% JFH expected to be around
expected to be around $\sim\SI{2e4}{\kelvin}$ and $\gamma \sim 1$ \citep{Aloisio_2019}; as time proceeds, $T_0$ decreases adiabatically while $\gamma$ is expected to increase and asymptotically approach a value of $1.62$ \citep{hui_1997}. The dynamical time that it takes the low-density IGM to respond to temperature changes at the Jeans scale (i.e., the sound-crossing time) is the Hubble time \citep{Gnedin_1998}.  
Therefore, at the end of and after the epoch of reionization ($z = 5-6$), the gas with long cooling time-scale still retains useful thermal memory of reionization, providing an indirect but more observable probe to constrain the reionization history \citep{Miralda_1994, hui_1997, Upton_2016, onorbe_2017}.

Many previous studies have measured the thermal parameters of the IGM through various summary statistics of the \lya~forest, e.g.,close quasar pairs \citep{Rorai_Hennawi_2013, Rorai_Hennawi_2017}, wavelet amplitudes \citep{Tsum_2002MNRAS.332..367T,Tsum_2010ApJ...718..199L,Tsum_2012MNRAS.424.1723G, gaikwad_2020, wolfson2021wavelets}, average local curvature \citep{Tsum_2011MNRAS.410.1096B,  Tsum_2014MNRAS.441.1916B, gaikwad_2020}, 
%% JFH add close quasar pairs and cite papers by Rorai DONE
the power spectrum of transmitted flux \citep{Tsum_2001ApJ...557..519Z, Tsum_2009MNRAS.399L..39V,Tsum_2017JCAP...06..047Y, Tsum_2017PhRvD..96b3522I, boera_2019, gaikwad_2020, wolfson2021wavelets}, or the decomposition of the 
%% JFH It is poor writing style to lump all the references at the end here such that one can determine which reference is which technique. It also results in a huge block of references which is ugly. Instead put the references after you discuss each summary statistic or method, which is the correct way to reference.  DONE
forest \citep{Haehnelt_1998,Tsum_2000ApJ...534...41R, Tsum_2000ApJ...534...57B, Tsum_2000MNRAS.318..817S, Tsum_2001ApJ...562...52M, Tsum_2012ApJ...757L..30R,Tsum_2014MNRAS.438.2499B,  Tsum_2018ApJ...865...42H}. 
%% JFH Why is Molly's wavelet paper not being cited above?? DONE
In this work, we use the autocorrelation function of the \lya~forest flux, which is the Fourier transform of the power spectrum, from mock data over a redshift range from $z = 5.4$ to $z = 6.0$. We consider $T_0$ and $\gamma$ as thermal parameters and the mean transmission $\langle F \rangle$
%% JFH Yuo have not provided any explanation of the correspondence between <F> and Gamma_HI here. need one
%% more sentence.  DONE
as a further astrophysical parameter to take account of the amplitude of the assumed uniform ultra-violet background 
%% JFH ultra violet background should not be captial, only the abbreviation DONE 
(UVB, $\Gamma_{\ion{H}{I}}$) field for this paper. This is done by re-scaling the optical depths of~\lya, $\tau$, along the skewer such that $\langle e^{-\tau} \rangle = \langle F \rangle$, while $\tau = n_{\rm HI} \sigma_{{\rm Ly}\alpha} \propto  1 / \Gamma_{\rm HI}$. 
%IS IT OK?

The simulations we use for this work are hence according to a "paint-on" method for a tight temperature-density relationship in Equation~\ref{eq:temperature-density relation} that only considers density fluctuation.
The choice of the autocorrelation function as our summary statistic and the semi-numerical method for simulations are described in full detail in \citet{wolfson2023forecasting} \citepalias[hereafter][]{wolfson2023forecasting}, which serves as the foundation for this paper and provides a baseline for comparison. Caveats of assuming such a uniform UVB model are discussed in Appendix \ref{appdx:UVB}. However, we emphasize that this paper is centered around the machine learning methodology that can be applied to future studies.

Previous attempts to constrain the thermal evolution of IGM mainly require cosmological hydrodynamic simulations for parameter inference with presumed likelihood functions (i.e., Gaussian likelihood). The likelihood estimation naturally requires a statistical 
%% JFH statistical
model, %at every likelihood sample, 
%% JFH what is a likelikhood sample? Not the correct term. DELETED
which often comes from linear or nearest-neighbor (nearest-grid point, NGP hereinafter) interpolation of simulation outputs \citep{McDonald_2006, Tsum_2017PhRvD..96b3522I, boera_2019, gaikwad_2020, Gaikwad_2021, wolfson2023forecasting, Arya_2024JCAP...04..063A}. This way of model interpolation for simultaneously varying multiple parameters demands a fine grid of computationally expensive high-resolution hydrodynamical simulations, which can require tens of thousands of GPU-hours to properly resolve the IGM's small-scale physics where the thermal information is manifested. The computational challenge leads to the creation of cosmological emulators by generating fast surrogate models for high-resolution simulations over a broad range of parameter space.
 
As machine learning (ML) solutions have been proven to %enhance the efficiency and accuracy of pattern recognition, noise reduction, and parameter estimation 
be useful in solving highly nonlinear problems particularly with deep learning studies %\citep[see][for reviews of ML]{deep_learning_LeCun,Goodfellow-et-al-2016}
, they have opened up a new avenue for cosmological emulation \citep[see][for reviews of ML applications in cosmology]{ML_cos_2023,Huertas-Company_Lanusse_2023}. Emulating the IGM's thermal history from \lya~forest in the ML context was first performed by \citet{walther_2019}, who emulated the \lya~flux power spectrum for varying thermal parameters using Gaussian processes, while other emulators have long been used to study the physics of different observables \citep{emu_Heitmann_2009, emu_2015Kwan, emu_2015Liu, emu_2015Petri, emu_2019Jennings, emu_2019McClintock, emu_2019Zhai, Hennawi_2024}. Popular techniques for emulating the \lya~forest include using Taylor expansion or quadratic polynomial interpolation \citep{Matteo_2006, Bird_2011, Palanque_2013A&A...559A..85P, Palanque_2015JCAP...02..045P, Yche_Taylor_2017JCAP...06..047Y, Palanque_Taylor_2020JCAP...04..038P}, Gaussian Processes \citep{Bird_2019, Rogers_2019, Walther_2021JCAP...04..059W, Pedersen_2021JCAP...05..033P, Rogers_2021PhRvD.103d3526R, Fernandez_2022MNRAS.517.3200F, Bird_2023JCAP...10..037B}, and neural networks \citep{Huang_2021, Harrington_2022, Wang_thermal_2022, Nayak_2023, Molaro_2023MNRAS.521.1489M, Cabayol_nn_2023MNRAS.525.3499C, nasir2024deep, Maitra_2024}. For the first two methods, limitations in the number of cosmological simulations lead to a sensitive choice of representation when attempting to span a sufficiently wide parameter space that can be run in a reasonably long time. \citet{Rogers_2019} proposed a solution to employ Bayesian optimization of the training set from a Latin hypercube sampling for the Gaussian process scheme. However, training on optimized sample points is not always feasible for achieving accuracy—sometimes, we simply do not have the option to select ideal samples. This is why we opt for neural networks (NN), which can effectively utilize a small number of training points, regardless of how the samples are distributed. 

We have witnessed a notable increase in applications of NN in \lya~forest studies: 
to name a few, generating \lya~forest for large surveys using only N-body simulation with a convolutional neural network \citep{Harrington_2022}, reconstructing the underlying neutral hydrogen density from \lya~transmission flux with a neural network \citep{Huang_2021}. For the particular interest of this work, the superiority of NN in learning the thermal history of the IGM has been demonstrated in recent studies; for instance, predicting the IGM temperature for each pixel from \lya~transmission flux at  $z = 2 - 3$ with a convolutional neural network \citep{Wang_thermal_2022}, using \lya~transmission power spectrum and transmission PDF to obtain the IGM thermal parameters at $z = 2.2$ with a residual convolutional neural network \citep{Nayak_2023}, predicting the \lya~optical depth-weighted density or the IGM temperature for each pixel from the \lya~transmission flux at $z = 4-5$ using Bayesian neural networks \citep{nasir2024deep}, and extracting the IGM thermal parameters at $z = 2-4$ from \lya~1D-transmitted flux in Fourier space using information maximising Bayesian neural network \citep{Maitra_2024}. However, a deep learning framework for modeling the thermal evolution of the IGM at higher redshifts is yet to be attempted.

%% JFH While this intro paragraph on ML applications in the Ly-a forest is nice, you have not really provided a very good introduction or exhaustive introduction to the use of emulators in the Ly-a forest, and why we need another one? What is wrong with Bird's approach? BC THEY ARE z = 2.2~4.2 N SENSITIVE TO TRAINING CHOICE? You did not describe the Taylor expansion emulators.  So your intro provides intro for everyhting except the problem we are actually interested in;(  REWRTE DONE

%% JFH You probably also need to cite this paper: DONE (pair w bird's paper) https://ui.adsabs.harvard.edu/abs/2019JCAP...02..031R/abstract 

%% JFH Why is the Hennawi et al. on IGM damping wings not be cited anywhere in this paper, since that is where the inference test and coverage testing is described in a ton of detail? IT WASNT PUB AT THE TIME I WROTE， DONE
This work thus presents a feedforward neural network \citep{Goodfellow-et-al-2016} emulator on only a small number of hydrodynamic simulations, which allows us to make precise thermal state inference from the IGM at $5.4 \le z \le 6.0$ without the aforementioned concerns. Even for a single mock \lya~data ($33$ cMpc h$^{-1}$ at $z = 5.4$ in this paper) with reasonable observational noise, we can infer thermal parameters of the IGM at the given redshift bin. Inspired by \citet{Hennawi_2024}, which precisely measured the neutral fraction and quasar lifetime from IGM damping wings, we speed up the procedure with automatic differentiation environment \texttt{JAX} and  Hamiltonian Monte Carlo (HMC) algorithm for sampling probability distributions. Measurements are compared to those from traditional NGP with Markov Chain Monte Carlo (MCMC) approach \citepalias{wolfson2023forecasting}. Finally, by performing statistical inference \citep{Hennawi_2024} on 100 realistic mock \lya~autocorrelation function data, we attempt to show that our posterior constraints of $T_0, \gamma, \langle F \rangle$ generated by the emulator are unbiased.

This paper is organized as follows. Section \ref{sec:simulation} summarizes the hydrodynamic simulations and the modeling of the \lya~autocorrelation function. Section \ref{sec:NN} presents the construction of the  emulator and its performance. In Section \ref{sec:inference}, we show how to use HMC to measure the thermal state of the IGM at each redshift with the emulator while incorporating emulation uncertainties. Section \ref{sec:inf_test} tests the statistical robustness of our inference process by running an inference test. We conclude in Section \ref{sec:conclusion} with a comprehensive summary of the advantages of our emulator and discuss potential future applications of our framework.

\section{Simulations and Models}\label{sec:simulation}
\subsection{Hydrodynamical Simulations and Forward Modeling}\label{subsec:sims}

\begin{figure}
    \includegraphics[width=\columnwidth]{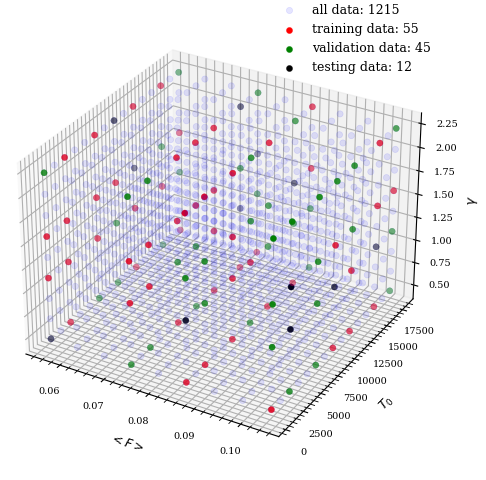}
    \caption{Data split in thermal parameter grid of $[T_0, \gamma, \langle F \rangle]$ for training,  test, and validation data at $z = 5.4$. Both the training and validation sets are kept at same relative positions in parameter space for each redshift $z = 5.4 - 6.0$, whereas the test data set has more points for  $z = 5.9 - 6.0$, further detailed in Appendix~\ref{sampling}.}
    \label{fig:train data}  
\end{figure}

This work uses a simulation box of size $L_{\text{box}} = 100$ comoving Mpc (cMpc) h$^{-1}$ run with $4096^3$ resolution in both dark matter and baryons with \texttt{Nyx} code \citep{almgren_2013}, a N-body and Eulerian hydrodynamical simulation code designed to simulate the Ly$\alpha$ forest. Simulation as such %would 
%% JFH Why is this hypothetical? Would take? It simply takes that long
takes about 16,000 GPU-hours each run. Details of the thermal model generation and simulation box 
%% JFH only the box details? ADD MODEL GEN
setting can be found in \citetalias{wolfson2023forecasting} Section 2.1. 
%% JFH Don't use the alias Paper 1. This is not a multi-part series. Refer to it as W23 or something like that. It is not paper 1, because this is not paper 2. DONE-W2023
However, the most important details will be summarized in the following.
We consider models in 
%% JFH at --> in 
seven redshift bins: $5.4 \leq z \leq 6.0$ with $\Delta z = 0.1$. 
To study different thermal states of the IGM, we generate the temperature of each cell following Equation \eqref{eq:temperature-density relation} with different values of $T_0$ and $\gamma$ for all densities. We sample 15 values of $T_0$ logarithmically and nine values of $\gamma$ linearly.  
Meanwhile, by re-scaling the optical depth such that the mean transmitted flux, $\langle F \rangle$, over all simulation skewers is a designated model value, nine linearly-spaced values of $\langle F \rangle$ are used to model a variety of potential $\langle \Gamma_{\ion{H}{I}} \rangle$ values. 
%% JFH you never explained the physics behind thees <F> \Gamma_HI connectino. See my comment above. DONE
To construct the grid, we choose central ``true'' values of $T_0~\text{and}~\gamma$ from a model similar to \citet{Upton_2016}, and $\langle F \rangle$ is centered on the values from \citet{bosman_2021_data} (all central values are presented in Table~\ref{tab:central vals}). At all $z$, we use the errors on the measurements reported in \citet{gaikwad_2020} at $z = 5.8$ ($\Delta T_0 = \SI{2200}{\kelvin}$ and $\Delta \gamma = 0.22$) 
%% JFH I'm not following how you used Gaikwad. Why are we assuming his measurements are right for our emulator? Plus the blue points span a much large range than you are saying in T0, i.e. from 0 -17500 DONE: more details on grid
and the redshift dependent $\Delta \langle F \rangle$ reported in \citet{bosman_2021_data} to construct the grid around the central ``true'' values at each redshift, i.e., modeling from $T_0 - 4\Delta T_0$ to $T_0 + 4\Delta T_0$, $\gamma - 4\Delta \gamma$ to $\gamma + 4\Delta \gamma$, and $\langle F \rangle - 4\Delta \langle F \rangle$ to $\langle F \rangle + 4\Delta \langle F \rangle$ in linear bins. This construction of the simulation grid results in a total of 1215 different thermal models at each $z$, shown as light blue points in Figure~\ref{fig:train data}, which is the parameter grid at $z = 5.4$, where central point of 
%% JFH central point of the grid DONE
the grid has $T_0 =\SI{9149}{\kelvin} , \gamma = 1.352, \langle F \rangle = 0.0801 $. 

To mimic the high-resolution observational data in real life, the skewers from each simulation box are forward modelled with a resolution of $R = \SI{30000}{}$ and a signal-to-noise ratio per \SI{10}{\kilo\meter\per\second} pixel ($\snr$) of $\snr = 30$ at all $z$. By adding flux-independent Gaussian random noise that is generated from the same random number seed to 1000 skewers 
%% JFH consistent is not the right word. You mean to say you kept the random number seed the same so the noise realization is always the same. Make that clear. DONE
in every simulation box at all $z$, we prevent introduction of further stochasticity in parameter inference. Considering the long box size of 100 cMpc h$^{-1}$, we split the skewers to two parts each of length $\Delta z = 0.1$, resulting in a total of 2000 forward-modelled skewers for each model. Details of the forward modeling can be read in \citetalias{wolfson2023forecasting} Section 2.2.

\begin{table}
    \centering
    \caption{
        This table lists the central ``true'' values of the redshift-dependent thermal state models used in this work.
        % include the range somehow?
        The last column states the central ``true'' value of $\langle F \rangle$ modelled in this work, which are the measurements from \citet{bosman_2021_data}. 
    }
	\label{tab:central vals}
    \begin{tabular}{|c|c|c|c|}
        \hline $z$ & $T_0$ (K) & $\gamma$ & $\langle F \rangle$ \\ \hline
        5.4 & 9,149     & 1.352    & 0.0801   \\
        5.5 & 9,354     & 1.338    & 0.0591   \\
        5.6 & 9,572     & 1.324    & 0.0447   \\
        5.7 & 9,804     & 1.309    & 0.0256   \\
        5.8 & 10,050    & 1.294    & 0.0172   \\
        5.9 & 10,320    & 1.278    & 0.0114   \\
        6.0 & 10,600    & 1.262    & 0.0089   \\ \hline
    \end{tabular}
\end{table}

\subsection{autocorrelation Function Models} \label{corr}
Instead of studying the Ly$\alpha$ forest flux itself, we investigate the autocorrelation function of it, which is calculated from:

\begin{equation}
    \xi_F (\Delta v) = \langle F(v) F(v + \Delta v) \rangle
	\label{eq:autocorr}
\end{equation}

where $F(v)$ is the flux of the Ly$\alpha$ forest and $v$ is the recessional velocity which we use for scales in autocorrelation function. By taking the expected value over all pairs of pixels with the same velocity lag, $\Delta v$, the uncorrelated white noise is centered around zero and thus unbiased. 
%% JFH it is not filtered out. It averages out and should be centered around zero, so unbiased. DONE
%Complicated spectral masking is also no longer needed for the removal of metal lines from other $z$ and \lya~damping wings from over-dense regions of \ion{H}{I}, see \citet{walther_2019}. 
%% JFH this is confusing and incorrect. Masking is still needed, it just has no impact on the statistic. Please be careful with what you are saying! OK-DELETED
It retains the same information as the power spectrum of the transmitted flux because it can also be computed via the Fourier transform of the dimensionless power spectrum.

The velocity bins in definition are set up as follows. With the smallest bin set at the resolution length $\SI{10}{\kilo\meter\per\second}$, a bin size of $\SI{10}{\kilo\meter\per\second}$ is added linearly up to $\SI{300}{\kilo\meter\per\second}$. Then logarithmic bin widths of $\log(\Delta v) = 0.029$ are applied to a maximal distance of $\SI{2700}{\kilo\meter\per\second}$, resulting in 59 velocity bins considered where the first 28 have linear spacing \citepalias{wolfson2023forecasting}. The smallest scales are computed with linear bins because they contain the most thermal information of \lya~flux 
%% thermal info of Ly-a flux is incorrect english
comparing to large scales. 

The \textit{mean model} of the autocorrelation function for each combination of thermal parameters is what we use to train the emulator. It is obtained by taking the average of the autocorrelation function over all 2000 forward-modelled skewers. The resulting correlation function models for different thermal states at $z = 5.4$ are plotted in figure 4 of \citetalias{wolfson2023forecasting}. 
%% JFH Citing the wrong figure. Use \label and \ref to avoid making such mistakes. ITS MOLLY'S PAPER

%% JFH It is werid that you are citing the figures out of order. If you need to cite Figure 4 now, you should make it Figure 2. By the way, you are citing the wrong Figure here. 

%% JFH I'm confused???? For computing the mean we don't need forward modelled skewers unless you are mocking up continuum error? In other words, you need to treat resolution effects, but you don't need to add noise since the noise averages to zero. This likely made it much harder to emulate!!

With the space-filling representation, visualized in Figure~\ref{fig:train data}, we use only 112 (132 for $z = 5.0-6.0$) out of the
1215 thermal models available at each $z$ for training and inference runs, 
less than 10\% of the data needed in the NGP interpolation method in \citetalias{wolfson2023forecasting}. This is necessary because for noisier and higher-resolution models, the number of cosmological simulations will be much fewer comparing to our forward modeling here. As indicated in the figure legend, we use a training set consisting of 55 models, a validation set of 45 models, and a test set of 12 models (32 models for $z = 5.9-6.0$). We conducted training experiments using data sets of varying sizes and determined that the chosen data set size represents the minimum required to achieve near percent-level accuracy across all redshifts. Increasing the size of the data set beyond this point yields only marginal improvements. Data sampling for training, validation, and tests 
%% JFH test --> tests
are 
%% JFH is described --> are described
described in full detail in Appendix~\ref{sampling}.

\subsection{Model Covariance Matrix} \label{sec: covar}
We draw \textit{a mock data set} by 
%% JFH from randomly --> by randomly 
randomly selecting and averaging the autocorrelation function over 20 forward-modelled skewers (i.e. 20 quasar sightlines) from the initial 2000 skewers from each simulation box. We then compute the covariance matrix for each thermal model from those mock draws:
\begin{equation}
    \Sigma_{\text{data}}(T_0, \gamma,\langle F \rangle) = \frac{1}{N_{\text{mocks}}} \sum_{i=1}^{N_{\text{mocks}}}(\boldsymbol{\xi}_i - \boldsymbol{\xi_\text{model}})(\boldsymbol{\xi}_i - \boldsymbol{\xi_\text{model}})^{\text{T}}
    \label{eq:covariance}
\end{equation}
where $\boldsymbol{\xi}_i =  \boldsymbol{\xi}_i(T_0, \gamma,\langle F \rangle)$ is the $i$-th mock autocorrelation function, $\boldsymbol{\xi_\text{model}} = \boldsymbol{\xi_\text{model}}(T_0, \gamma,\langle F \rangle)$ is the model value of the autocorrelation function, and $N_{\text{mocks}}$ is the number of mock data sets used, which is set to 500,000 for all models and redshifts in this work. This number of mock draws has been shown to be sufficient for minimizing calculation errors in covariance matrices, as demonstrated in appendix B of \citetalias{wolfson2023forecasting}. Note that $\Sigma_{\text{data}}(T_0, \gamma,\langle F \rangle)$ is calculated for each combination of thermal parameters, resulting in 1215 separate computations for each $z$. To better visualize the correlation of the data error at all scales, a correlation matrix, $C$, is defined to standardize across rows and columns around the diagonal elements.

\begin{equation}
        C_{jk} = \frac{\Sigma_{jk}}{\sqrt{\Sigma_{jj}\Sigma_{kk}}}.
        \label{eq:correlation}
\end{equation}

\begin{figure}
	\includegraphics[width=\columnwidth]{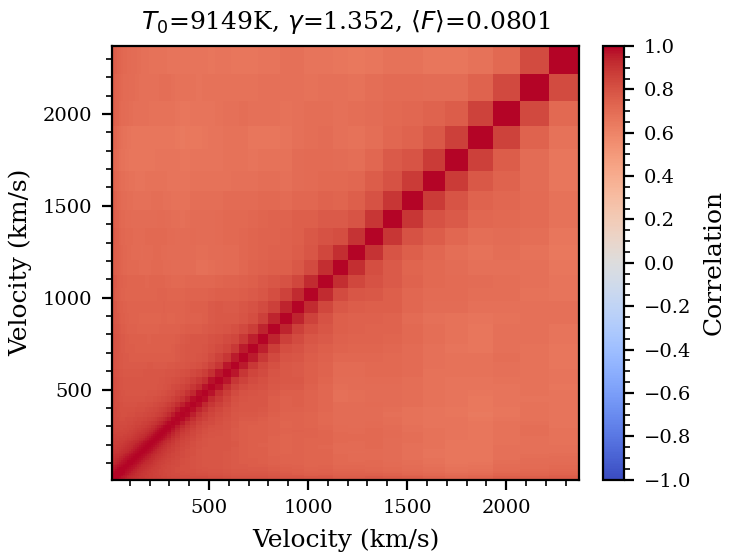}
    \caption{
        Correlation matrix at $T_0 = \SI{9149}{\kelvin}$, $\gamma = 0.0801$, $\langle F \rangle = 1.352$ for $z = 5.4$.  This illustrates that all bins in the autocorrelation function are highly correlated with each other.
    }
    \label{fig:covar_data}
\end{figure}

An example of the correlation matrix for the model at the center of $z = 5.4$ parameter grid, where $T_0 = \SI{10611}{\kelvin}$, $\gamma = 0.0591$, $\langle F \rangle = 1.338$, is shown in Figure \ref{fig:covar_data}. We can see that all velocity bins are highly correlated because the autocorrelation function at each bin is calculated from the same pixels of $F(v)$.

%% JFH Comment on the fact that the correlation function is highly correlated and briefly explain why. DONE

\section{JAX-Neural Network Emulator} \label{sec:NN}
%% JFH This subsection is too long. I would dramatically shorten it. This is not a paper advertising Jax. OK

We train a fully-connected neural network (NN) emulator to learn the dependence of the \lya~forest autocorrelation
%% JFH ly-a forest autocorrelation 
function on the thermal parameters. While only trained on a small number of thermal models, the emulator interpolates at any place in a larger parameter space. Consequently, it reduces the need for computationally expensive simulations to accurately interpolate in parameter space. It also speeds up measurements of physics-informed parameters from the observable through function compilations with \texttt{JAX}, a High-Performance Array Computing library, which will be described 
%% JFH introduced --> described
in Section~\ref{jax}. The architecture of our NN is described in Section~\ref{mlp}. We present the emulation accuracy and the fit to the test data set in Section~\ref{sec: performance}.

\subsection{\texttt{JAX}: Auto-Differentiation and Just-in-Time Compilation \textsc{python} Library}\label{jax}
%The aim of this section is to have an overall introduction of the powerful functionality of \texttt{JAX} \citep{jax2018github} and thus the reasons why we adopted this \textsc{python} library to build our NN emulator instead of others (TensorFlow, PyTorch, etc.).

The most important reason why we build our NN in \texttt{JAX} is its automatic differentiation nature. It significantly speeds up traditional methods for computing derivatives in likelihood estimation. %The traditional methods of computing derivatives are either estimation in infinitesimal differences or analytical differentiation from formulas. Both methods have particular requirements on the function to maintain a certain degree of accuracy. 
While taking the derivatives of native \textsc{python} and \texttt{Numpy} code, we can handle a large subset of \textsc{python}'s features, including loops, ifs, recursion and closures, and we can even take derivatives of derivatives of derivatives \citep{jax2018github}. %The easy-to-use \texttt{NumPy}-inspired interface can also convert existing \texttt{NumPy} code and run these calls on the accelerator if a GPU (or TPU) exists, which has the potential to be much faster than on CPU. 
Despite the slower compilation of \textsc{python} compared to other compiled languages, \texttt{JAX} can compile functions end-to-end with Just-in-Time compilation (\texttt{jit}) using XLA (Accelerated Linear Algebra). It allows compilations for both CPU and accelerators (GPU/TPU), gaining orders of magnitude speedup in \textsc{python}.

%The other reason \texttt{JAX} excels other
%% JFH fix english in this sentence DONE
%machine learning frameworks empowered with function compilation is the easy implication 
%%JFH implication is not the right word here. 
%of its \texttt{NumPy}-inspired interface. 
%Another reason JAX outperforms other machine learning frameworks with function compilation capabilities is its easy-to-use NumPy-inspired interface. The existing \texttt{NumPy} code can be directly converted to \texttt{JAX} with \texttt{jax.numpy}. If a GPU (or TPU) exists, these calls of \texttt{NumPy} code run on the accelerator and have the potential to be much faster than on CPU. 

Lastly, \texttt{JAX} is featured with its vectorizing map transformation, which automatically transforms the parameters of a function to vectors while maintaining the element-wise function operation. Vectorization of both inputs and outputs of a function allows iteration over designated inputs without explicitly writing the loop. In general, \texttt{JAX} allows us to write the NN library in plain \textsc{python} code and still benefit from the speed of compiled code and tackles the onerous task of computing the gradient for a complex model required by HMC \citep{hoffman2014no}, as discussed later.

\subsection{Neural Network Emulator Architecture}\label{mlp}

\begin{figure*}
    \centering
	\includegraphics[width=2.2\columnwidth]{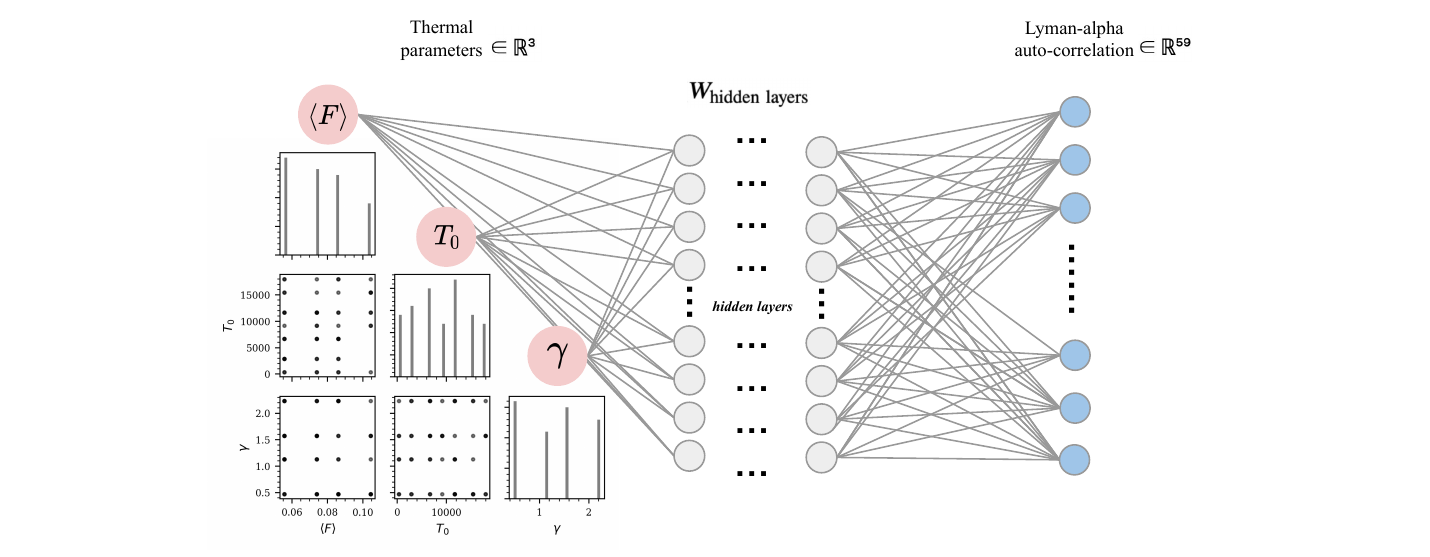}
    \caption{MLP architecture of NN emulator at $z = 5.4$. The NN learns to map from the input thermal parameters of $ \mathbb{R}^3$ (left-hand corner plot is the parameter distribution for training) to the corresponding output \lya~autocorrelation function of $ \mathbb{R}^{59}$ through a fully-connected multi-layer structure. Each epoch repeats the learning process through all layers and computes the loss from the loss function and the weights and biases ($W_{\mathrm{hidden~layers}}$ hereinafter) from all the neurons. By marginalizing over the loss, an optimizer updates $W_{\mathrm{hidden~layers}}$ for each epoch and eventually finds the minimal loss with the best trained $W_{\mathrm{hidden~layers}}$, that can be used to map any set of thermal parameters to a \lya~autocorrelation function in the same parameter space as the training data (i.e., at the same redshift).}
    \label{fig:MLP}
\end{figure*}

Multi-Layer Perceptrons (MLP) are supervised learning algorithms designed to understand complex relationships $f(.) : \mathbb{R}^n \xrightarrow{} \mathbb{R}^m $ \citep{MLP}. MLP is the standard structure for a neural network, which consists of fully interconnected neurons through layers. This structure, illustrated in Figure \ref{fig:MLP}, allows full grasp of nonlinear functions and hence data relationships%regardless of the dimension change. 
, even when the input and output have drastically different dimensions.
%% JFH dimension change? CHNGED
We adopt this structure to emulate the model values of the \lya~autocorrelation function described in Section~\ref{corr} from combinations of thermal parameters. 

With space-filling representation for training, validation, and test, as described in full in Appendix~\ref{sampling}, we train the emulator separately for each redshift. The input space dimension is therefore $n = 3$ for $[T_0, \gamma, \langle F \rangle]$, and the output space dimension is $m = 59$ corresponding to the number of velocity bins of each $\xi_F$. This project makes use of \texttt{Haiku} \citep{haiku2020github}, a simple neural network library for \texttt{JAX}, for composing modules defined in the usual networks to construct our custom MLP architecture.

The conventional loss function in NNs is mean square error (MSE). However, as we want to handle on the different scaling of small and large velocity bins, we chose the Relative Mean Absolute Error (RMAE) in physical units as our loss function since it exclusively tells the relative errors on various scale of data across velocity lags. 

\begin{equation}
    \mathrm{RMAE} = \frac{1}{m}\sum_{i = 1}^{i =m}\left | \frac{ \xi_{\mathrm{model},i} - \xi_{\mathrm{NN},i}}{\xi_{\mathrm{model},i}}\right |
    \label{loss}
\end{equation}
where $m$ is the dimension of data,  $\xi_{\mathrm{model},i}$ is the correlation model value at velocity bin $i$, and $\xi_{\mathrm{NN},i}$ is the predicted correlation function at velocity bin $i$. We also test our emulator on other loss functions, detailed further in Appendix~\ref{hparam}.

We use Adamw optimizer from \texttt{Optax} \citep{deepmind2020jax} to update the weights and biases for all neurons. It uses weight decay to regularize learning towards small weights. Through iterative epochs, our NN learns to emulate the autocorrelation function from the three thermal parameters $[T_0, \gamma, \langle F \rangle]$ with over-fitting prevention (detailed settings in Appendix~\ref{hparam}).

We experimented with different network architectures to optimize performance by marginalizing the validation loss with \texttt{Optuna} \citep{akiba2019optuna}, an automatic hyperparameter optimization software framework particularly designed for machine learning. In the end, we are able to construct our emulator for each $z$ with separate optimal choices of hyperparameters, reported in Appendix~\ref{hparam}.

\subsection{Performance evaluation: emulation accuracy}\label{sec: performance}
We use the test data set to test the performance of our emulator. These 12 (32 for $z = 5.9-6.0$) models were never used during the training of the emulator, serving as a proxy for new \lya~data.  

\begin{figure}
	\includegraphics[width=\columnwidth]{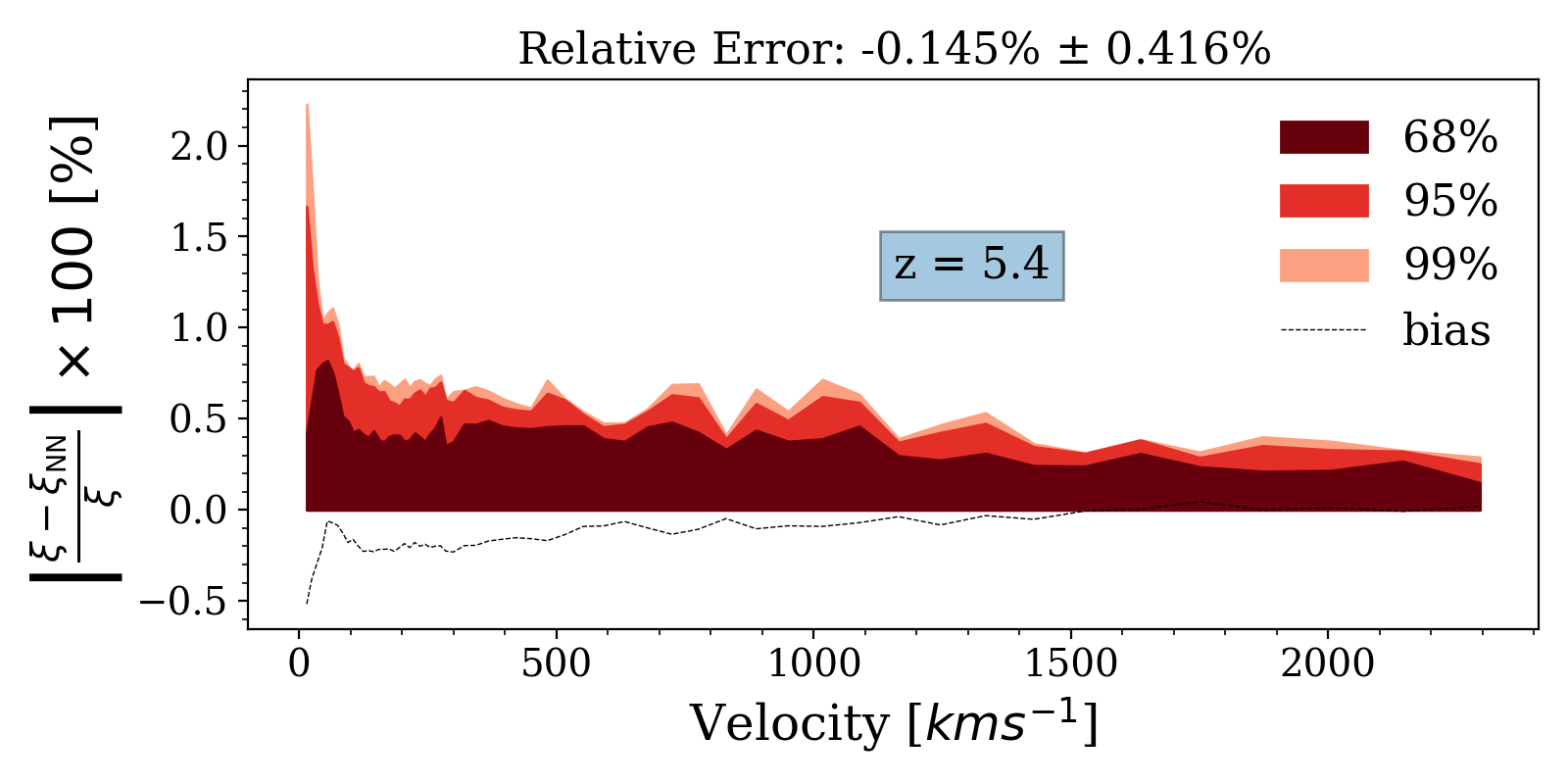}
    \caption{Emulation error for $z = 5.4$. It shows mean (dotted line) and standard deviation ($68\%$ percentile contour) of the relative percentage error evaluated from the 12 \lya~test data set. $68\%$ of the percentage errors of the test data set are restrained within 1\%.}
    \label{fig:accuracy}
\end{figure}

%% JFH Is the shading in Figure 4 necessary? It makes it hard to see the the bias? The aspect ratio for this figure is also werid, I would make the y-axis direction a bit larger to make it easier to see. 
The metric we used to evaluate the emulation accuracy is the relative percent absolute error, labeled on the y-axis of Figure~\ref{fig:accuracy}, consistent with the loss function not averaged across all velocity bins. 
%% JFH You might comment on the fact that is also basically your loss, i.e. your loss is this averaged over all the velocity bins. DONE
It reports the accuracy on the 12 test data sets at redshift $z = 5.4$ in the 68th, 95th, and 99th percentiles, as indicated in the legend. 99\% of the the test \lya~autocorrelation functions have error under $3.0\%$, with the overall average error of $0.145\% \pm 0.416\%$ across the velocity bins. From the figure, we can see that the error is typically worst on smaller scales. This is due to the increased non-linearity at small scales, characterized by a more rapidly changing shape resulting from higher signals and smaller bin sizes,
%% JFH Why should higher signal and smaller bins make it harder to emulate? I think the point is that the shape changes more steeply with the parameters, making it harder to model with a fixed set of NN weights. DONE: ADD SHAPE CHNG
making them inherently harder to model accurately. Appendix~\ref{appdix:other_z} presents the consistently strong emulation performance at other redshift.
%Figure~\ref{fig:test_overplot} presents the emulation fit results for nine randomly selected models from the test set. The results demonstrate a good agreement between the emulation and the space-filling models.

%% JFH English is poor in this last sentence "consistently well" DONE
\begin{comment}
\begin{figure*}
    \centering\includegraphics[width=2\columnwidth]{figures/performance/test_overplot_z54_train_55_bin59_seed_11_mape_l2_0_perc_True_activation_tanh.png}
    \caption{Emulation predictions for 9 test models at $z = 5.4$. (Upper panels) Solid red lines represent the \lya~autocorrelation function model values, while dashed blue lines are the NN emulator predictions. (Lower panels) Black lines are the percentage residuals of the emulation with the $\pm 1.0\%$ in red shaded region. }
    \label{fig:test_overplot}
\end{figure*}
\end{comment}

\section{Thermal Parameter Inference} \label{sec:inference}
The NN emulator allows us to interpolate in the parameter space of $T_0$, $\gamma$, and $\langle F \rangle$ and map them onto the autocorrelation function of the Ly$\alpha$ forest on and off the 
%% JFH training parameter grid DONE
training parameter grid without running 
%% JFH new DONE
new simulations. This allows us to constrain the thermal state of the IGM with the observational or mock Ly$\alpha$ forest flux at a given $z$. In order to do this inference, we need to sample the likelihood for each thermal model. We chose the Hamiltonian Monte Carlo (HMC) as our sampling algorithm while employing the multi-variate Gaussian likelihood with modification to propagate the emulator error. 

HMC \citep{duane_1987} is an advanced, gradient-based extension of the traditional Markov Chain Monte Carlo (MCMC) algorithm designed for efficient sampling in high-dimensional parameter spaces. Unlike regular MCMC, HMC leverages gradient information to navigate the posterior landscape, favoring regions with higher posterior probabilities and producing more efficient sampling trajectories. When combined with our differentiable emulator, HMC further accelerates the inference process significantly.

\subsection{Parameter Estimation: Bayesian Inference} 
The log-posterior distribution from which we sampled parameters follows  Bayesian inference. 
%% JFH I've never heard of "gradient based Bayesian inference" What is it? DONE: DEL
This approach relates the posterior distribution $p(\boldsymbol{\theta}|\mathbf{d})$ to the likelihood function $p(\mathbf{d}|\boldsymbol{\theta})$ according to Bayes' theorem:

\begin{equation}
\label{bayesian}
    p(\boldsymbol{\theta}|\mathbf{d}) = \frac{p(\mathbf{d}|\boldsymbol{\theta})p(\boldsymbol{\theta})}{p(\mathbf{d})}
\end{equation}

where $p(\boldsymbol{\theta})$ stands for the prior knowledge of the parameters, and $p(\mathbf{d})$ is a normalization factor commonly ignored during inference. Parameters we infer here are thermal parameters, i.e., $\boldsymbol{\theta} = [T_0, \gamma, \langle F \rangle]$, and the input is the autocorrelation function of any \lya~forest flux at given redshift, i.e., $\mathbf{d} =\boldsymbol{\xi}$. In this context, we use $\boldsymbol{\xi}$ to denote the correlation function of the flux itself, distinct from the relative flux fluctuation, to avoid potential ambiguity. We used non-informative %flat 
%% JFH uninformative
priors for all the parameters. Since the emulator was trained on fixed ranges of parameters, a transformation of a bounded parameter vector $\boldsymbol{\theta}$ into an unbounded parameter vector $\mathbf{x}$ using a logit transformation was used. 
%% JFH Next sentence is a computing detail that can be omitted! OK
%By taking advantege of \texttt{jax.vmap} to vectorize the transformation, it can be done on either a single parameter vector or a batch of parameter vectors from HMC sampling.
The likelihood function $p(\mathbf{d}|\boldsymbol{\theta})$ in this implementation is a multi-variate Gaussian likelihood $\mathcal{L}(\boldsymbol{\xi}|\boldsymbol{\theta})$: 

\begin{equation}
        \mathcal{L} = \frac{1}{\sqrt{\det(\Sigma) (2 \pi)^{n}}} \exp \left( -\frac{1}{2} (\boldsymbol{\xi} - \boldsymbol{\xi_\text{NN}(\boldsymbol{\theta})})^{\text{T}} \Sigma_\text{data}^{-1} (\boldsymbol{\xi} - \boldsymbol{\xi_\text{NN}(\boldsymbol{\theta})}) \right) 
        \label{eq:gauss_like}
\end{equation}

where $\boldsymbol{\xi_\text{NN}}$ is the assumed error-free model value of the autocorrelation function predicted by the emulator, evaluated at any thermal parameters $\boldsymbol{\theta}$, $\Sigma_\text{data}$ is the model-dependent covariance matrix of the data $\boldsymbol{\xi}$ estimated by Equation \eqref{eq:covariance}, and $n=59$ is the number of bins in the autocorrelation function. 
Note that we fixed the covariance matrix for the interpolation process under the assumption that, for any given data, its covariance can be estimated from the data itself. %and does not strongly evolve as a function of thermal parameters. 
%% JFH The last sentence is incorrect and unnecessary. We are not assuming this and it is incorrect to assume this. Instead we are assuming we can measure the covariance frmo the data. 
For real-life observational data, both the data covariance and the correlation function are hence required as inputs for our thermal parameter inference. This likelihood is assuming that the \lya~autocorrelation function at this redshift range is Gaussian distributed about the mean for each bin, but this is an incorrect assumption for low $\langle F \rangle$, which will fail the inference test later on, as discussed in Appendix~\ref{appdix: gaussian_data} and \citetalias{wolfson2023forecasting}.

%% JFH add caveats on assuming a Gaussian likelihood here, according to Molly's paper. THIS IS ASSUMING THE DATA IS GAUSSIAN, BUT IT'S NOT, DISCUSSED IN APPENDIX.

%% JFH It is poor style to use 's in a title. DEL
\subsection{Neural Net Error Propagation}\label{sec:nn err}
\begin{figure*}
    \includegraphics[width=\columnwidth]{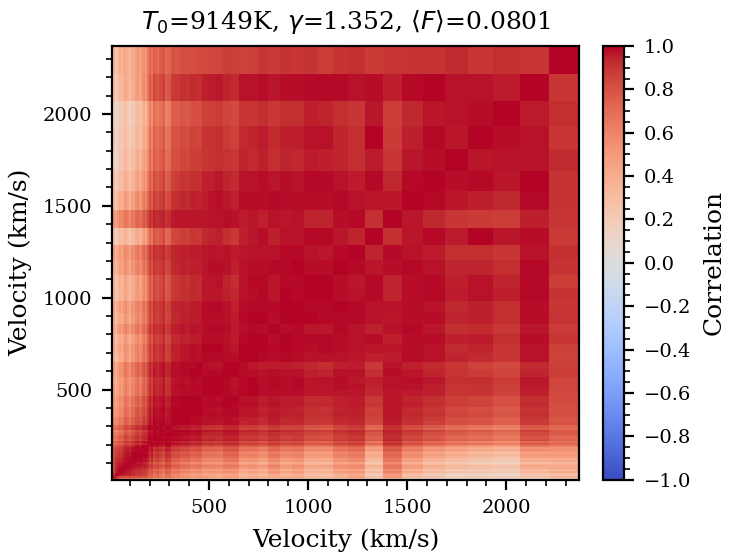}
    \includegraphics[width=0.95\columnwidth]{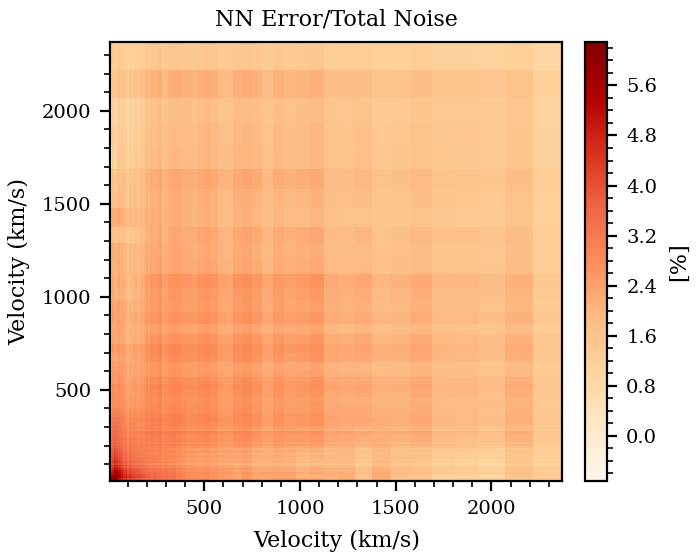}
    \caption{(Left) Correlation matrix of the estimated NN emulator's prediction error at $z = 5.4$. (Right) Percentage fraction of NN prediction error to the total error for inference as defined in Equation \eqref{eq:cova_frac}. We can see that the prediction error is up to 6\% on small-scale off-diagonal elements, so it is significant in inferring the thermal information of the IGM.}
    \label{fig:err_prop}
\end{figure*}

The thermal parameters are inferred from the physical data of the Ly$\alpha$ forest whose noise is taken into account by adopting the covariance matrix from the simulation data. Higher precision of the data would improve the inference precision but, on the other hand, demands an equal increase in precision of theoretical models from the NN emulator. In other words, we should take the emulator error into account as the interpolation error, or else it would bias the inference results by underestimating the uncertainties.
%% JFH Need one more sentence here more clearly stating why the NN error budget needs to be taken into account. DONE: THE ERROR OF EMULATION WOULD AFFECT THE RESULTS... 
Hence, an approximation of the NN emulator's error is required to avoid biasing the parameter inference. \citet{Grand_n_2022} provide a Bayesian solution that aligns with our goal. Instead of the procedure suggested in \citet{Grand_n_2022} to sample a separate data set to estimate the error in the emulator prediction, we directly use the test data set because running more models to train this error is expensive, so we opt for only 12.
%% JFH What is "abundant model sampling"?. I think you need to just clearly say that running more models to train this error
%% is expensive so we opt for only 12!! DONE
We first compute the covariance matrix of the NN prediction error
%% JFH Please don't call this an error matrix. It is a covariance matrix!! Use the correct terminology. DONE
approximated from the 12 test data (32 for $z = 5.9-6.0$).
\begin{equation}
    \Sigma_\text{NN} = \frac{1}{T - 1} \sum^T_{t=1}(\mathbf{\Delta_\text{NN,t}}-\bar\Delta_\text{NN})(\mathbf{\Delta_\text{NN,t}}-\bar\Delta_\text{NN})^\intercal 
    \label{eq:sig_nn}
\end{equation}

where $t$ runs over the samples of the test set. Defining $\mathbf{\Delta_\text{NN,t}} = \boldsymbol{\xi_\text{NN,t}} - \boldsymbol{\xi_\text{NN,t}}$ to be the NN prediction error for each test sample and $\bar\Delta_\text{NN} = \langle \mathbf{\Delta_\text{NN,t}} \rangle$ to be the mean of NN error over all test data (corresponding to the bias in Figure~\ref{fig:accuracy}) we have the error covariance matrix. The error-propagated likelihood function hence is modified as 
%% JFH as follows. Also don't add a blank space between text and equations, as that starts a new paragraph and messes up the 
%% equation spacing. 
the following:
\begin{equation}\label{eq:nn error}
    \mathcal{P}(\boldsymbol{\xi}|\boldsymbol{\theta}) \propto \mathrm{exp} \left( - \frac{1}{2} | \Delta(\boldsymbol{\theta}) -\bar\Delta_\text{NN}|^T (\Sigma_\text{data}+\Sigma_\text{NN})^{-1}| \Delta(\boldsymbol{\theta})-\bar\Delta_\text{NN}| \right)
\end{equation}
%% JFH No whitespace here. Equations are part of the paragraph. DONE
where $\Delta(\boldsymbol{\theta}) = \boldsymbol{\xi} - \boldsymbol{\xi_\text{NN}(\boldsymbol{\theta})}$ as in Equation \eqref{eq:gauss_like}, $\Sigma_\text{data}$ is the model-dependent covariance matrix of the data from Equation \eqref{eq:covariance}, and $\Sigma_\text{NN}$ is from Equation \eqref{eq:sig_nn}. The Gaussianity condition of $\mathbf{\Delta_\text{NN}}$ is also tested using different seeds for generating the test set, as detailed in Appendix~\ref{sampling}. Note that when $\bar\Delta_\text{NN}$ goes up, Equation \eqref{eq:nn error} eliminates the NN prediction
%% JFH Don't use net's. It is bad style. DONE
bias via subtraction of the error, which ensures the emulator uncertainties do not yield biases in parameter inference. 

To demonstrate the importance of this procedure for propagating NN errors, Figure \ref{fig:err_prop} shows an example of $\Sigma_\text{NN}$ 
%% JFH make it clear you show the correlation matrix in left panel. DONE
at $z = 5.4$ in left panel and its fraction of the total uncertainty in right panel. We calculate the fraction through:
%% JFH no whitespace!
\begin{equation}
    \text{NN Error/Total Noise} = \frac{\Sigma_\text{NN}}{\sqrt{\text{diag}(\Sigma_\text{data}+\Sigma_\text{NN}) \otimes \text{diag}(\Sigma_\text{data}+\Sigma_\text{NN})}}
    \label{eq:cova_frac}
\end{equation}
%% JFH How about a comment somewhere on why the covariance of the emulation error is always positive, since we scratched our heads about this for quite sometime and it required that you explain it to me. 
Note that the error budget is more influenced by the emulator's error on smaller scales ($\leq \SI{200}{\kilo\meter\per\second}$), particularly essential to measuring the thermal information of the IGM
\footnotetext[1]{\label{bimodal}The NN's $\sim6\%$ contribution to the total uncertainty and off-diagonal correlations at smaller velocity bins solve the bimodality in parameters' posterior distributions for some mocks.}. 
%% JFH it is incorrect to say it is dominated by the emulatoin error. It is only 5%!!! This is incorrect. 5% is not dominating the error budget. DONE 
This is because of the following: while the error bars from the data shrink significantly in this regime due to larger signals and averaging over more bin pairs, the fractional accuracy of the emulator predictions in this regime is slightly lower, as Figure~\ref{fig:accuracy} illustrated. The off-diagonal elements of NN error fraction matrix display information in dependence between emulation in different velocity bins, potential to affect the inference result\footnote[2]{Note that the covariance of the emulation error is all positively correlated because we have the same hyperparameter setting to estimate all scales in a single $z$. Also, by the design of our NN, all predictions of \lya~autocorrelation function are smooth. Therefore, the prediction is either lower or higher than the target across all bins when you add all the test set}.
%% JFH If you are not going to show this bimodality thing, don't mention it. Or make it a footnote. DONE

A full inference procedure for a random mock \lya~autocorrelation function is described as follows. Using Equation \eqref{eq:nn error} as the likelihood function, $p(\mathbf{d}|\boldsymbol{\theta})$, the potential energy for Hamiltonian Monte Carlo (HMC) sampling is calculated. The sampler inputs the correlation data as $\xi$ each time for a random posterior draw $\xi_\text{NN}$ generated by the emulator. By marginalizing over individual posteriors, the measurement for each parameter is obtained. For HMC sampling, we employ the No-U-Turn Sampler (NUTS) \citep{hoffman2014no} implemented in \texttt{NumPyro} for HMC sampling, which provides significant computational efficiency, being orders of magnitude faster than other libraries \citep{phan2019composable}. Details of the procedure and sampler settings can be found in Appendix~\ref{appdix:HMC_setting}. 

Figure \ref{fig:fit} shows an example of implementing the inference procedure on a random forward-modelled mock autocorrelation function
%% JFH The "mean model" procedure has not been defined anywhere has it? You need to explain to the reader what this is!! ADDED ABOVE
(defined in  \ref{sec: covar}) of  $T_0 =\SI{9149}{\kelvin} , \gamma = 1.352, \langle F \rangle = 0.0801 $ at $z = 5.4$. The figure displays the resulting models from our inference procedure as follows: the faint blue lines represent random posterior draws from the HMC sampler, while the solid red line denotes the inferred model, obtained as the median of each parameter's samples determined independently via the 50th percentile of the HMC draws (faint blue lines). The green dashed line indicates the true model correlation for the simulation from which the mock data was generated. The black points correspond to the input mock data, with error bars derived from the diagonal elements of the model-dependent covariance matrix. Finally, the thick yellow line represents the emulation model with the maximal combined probability. 
%% JFH This is incorrect. You do not evaluate the model at the marginalized median of each parameter,which is what you are 
%% describing here. Instead standard practice is to take the median of the blue curves!!!!! CHNGED TO DRAWS
%% JFH I don't think showing the maximum likelihood model is necessary. 
The true values of $T_0$, $\gamma$, and $\langle F \rangle$ are displayed in green text, while the inferred measurements are annotated in red, with their associated uncertainties represented at the 68th percentile. The purple annotations indicate two fitting scores to the true mean model (green dashed line), where values approaching one signify a better fit. The close agreement with the mean model value of the autocorrelation function highlights the emulator's ability to accurately capture the underlying thermal model, even in the presence of observational or statistical noise from the random mock data. %Bottom panels of Figure \ref{fig:corner}
%% JFH Figure's does not work. No apostrophe's it is poor style for a figure reference. DONE
%show marginalised likelihood posteriors from the mean model value (left) and the random mock (right) with comparison in using the likelihood as in Equation \eqref{eq:nn error} that includes emulation error in orange lines and in Equation \eqref{eq:gauss_like} that does not in blue lines. 
The close fit to the mean model value of the autocorrelation function demonstrates that the emulator accurately captures the underlying thermal model in the presence of observational or statistical noise from the random mock data.

Figure \ref{fig:corner} shows the marginalized likelihood posteriors for the random mock (right) from Figure \ref{fig:fit} and its corresponding mean model (left), i.e. what we trained on, calculated as the average of all mocks with the same combination of parameters. The methods we obtained these data from simulations are defined in Section~\ref{corr} and \ref{sec: covar} respectively. Comparison of using the likelihood from Equation \eqref{eq:nn error}, which includes emulation error (orange lines), and Equation \eqref{eq:gauss_like}, which does not (blue lines) is also plotted.
%% JFH explain in words the different likelihoods, don't just cite the equation. explain that one includes emulation erorr and the other does not. DONE
Sampling across all the HMC chains provides thermal parameter measurements at the 50th percentile, with uncertainties derived from the 16th and 84th percentiles, as reported at the top of each 1D posterior. The deviation of parameter measurements from the true values (red lines) for the random mock data comparing to the mean model shows that the random noise of data still affects the inference results, and the constraining power of each parameter can depend on the luck of the draw when selecting the mocks. This is why it is necessary to conduct an inference test on a number of random mocks in the following section.

%% JFH This is a bit confusing since the top panel shows the mock, but then you show two sets of contours for the mean model and mock data. The caption title says you are doing both, but the top panel only is the mock. My suggestion is that you 
%% break this up into separate figres. DONE You never described this "mean model" procedure anywhere, so it is all kind of confusing. DONE: refer to appdix 
\begin{figure*}
    \includegraphics[width=2\columnwidth]{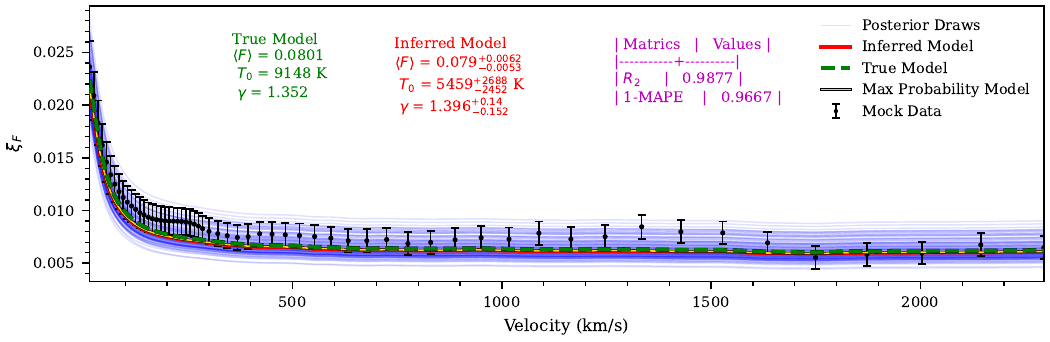}
    \caption{Emulation fit after our inference procedure applied to a random mock data set with parameter values of $T_0 =\SI{9149}{\kelvin} , \gamma = 1.352, \langle F \rangle = 0.0801$ at $z = 5.4$ as in Table~\ref{tab:central vals}. The faint blue lines represent random posterior draws from the HMC sampler. The solid red line indicates the inferred model derived from the median of each parameter’s samples, determined independently via the 50th percentile of the HMC chains. The dashed green line corresponds to the true model correlation for the simulation from which the random mock is drawn. The black points with error bars represent the input mock data, with errors derived from the diagonal elements of the model-dependent covariance matrix. Lastly, the thick yellow line shows the emulation model with the maximal combined probability.}
    \label{fig:fit}
\end{figure*}

\begin{figure*}
      \includegraphics[width=\columnwidth]{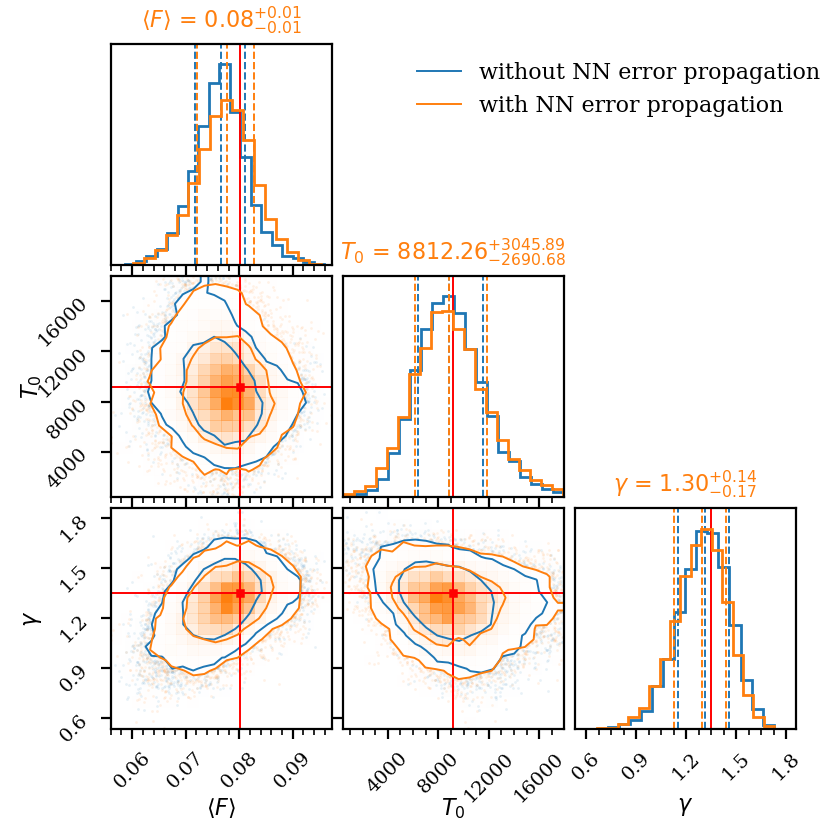}
    \includegraphics[width=\columnwidth]{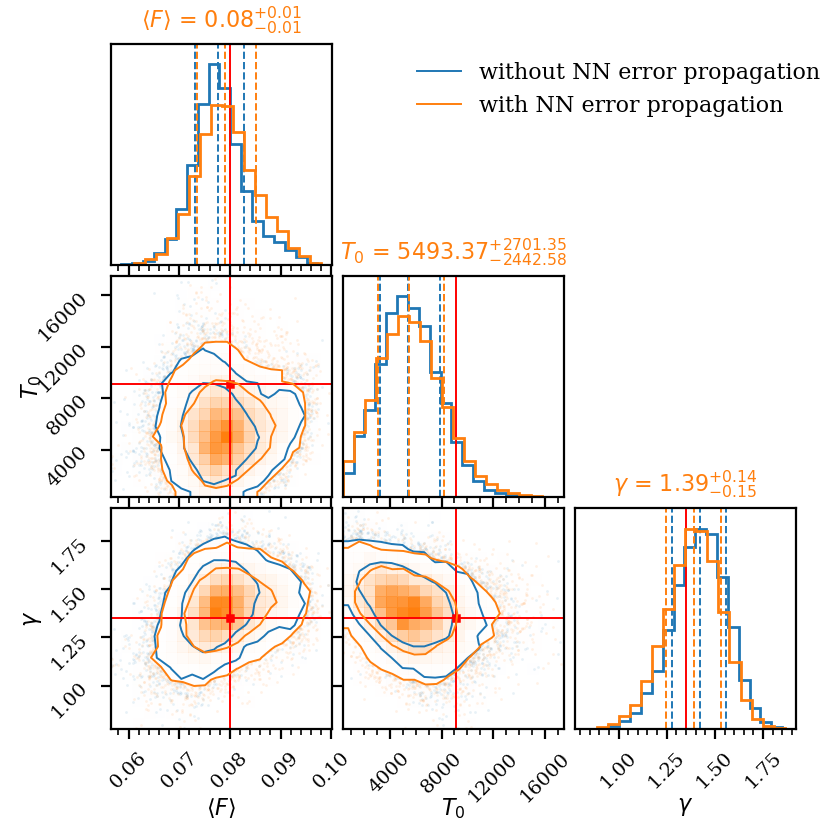}
        \caption{Marginalized likelihood posteriors with (orange) and without (blue) NN error propagation in likelihood for the parameter inference from the mean model (left) and a random mock data (right) with the same parameters of $T_0 =\SI{9149}{\kelvin} , \gamma = 1.352, \langle F \rangle = 0.0801$ at $z = 5.4$. The contours show 68\% and 95\% confidence intervals. The true parameter values are shown by red lines. Dash lines show values at 16th, 50th, and 84th percentiles, corresponding to the inference results for $T_0$, $\gamma$, and $\langle F \rangle$ at the top of each column.}
    \label{fig:corner}
\end{figure*}
As is evident from the figure, the orange posteriors are generally wider and have different shapes (see footnote 1) compared to the blue ones, proving the effect of error propagation, even when the emulator error is low across the full-scale range.
%% JFH Same thing here. Poor writing style to say Equation 10 effectively does something. Say it in words. OK
This could be explained by the additional term in the covariance that widened the posteriors in  Equation \eqref{eq:nn error}. 

%% JFH Left off here. 

\section{Results} \label{sec:results}
\subsection{Inference Test}\label{sec:inf_test}
To test the robustness of the inference procedure described in for all thermal models, an inference test was conducted on the coverage probabilities as the final credibility check of any assumption we made during HMC inference (i.e., likelihood function, priors, and parameter transformation). The formalism behind the inference test (i.e. coverage test) method is elaborated further in \citet{Hennawi_2024}, who assessed the reliability of its continuum-reconstruction PCA inference method for analyzing the IGM damping wings. The key steps are outlined below. The coverage probability $C(\alpha)$ of a posterior credibility level $\alpha$ is the fraction of the time in which the true parameters lie within the volume enclosed by the corresponding credibility contour in repetitions of the experiment. In our test, 100 random forward-modelled mock data sets were sampled uniformly from the flat priors on $T_0$, $\gamma$, and $\langle F \rangle$. To pass the inference test, the coverage probability should equal the posterior credibility for every level with one inference per mock for a total of 100 mock data sets.

\begin{figure}              
    \includegraphics[width=\columnwidth]{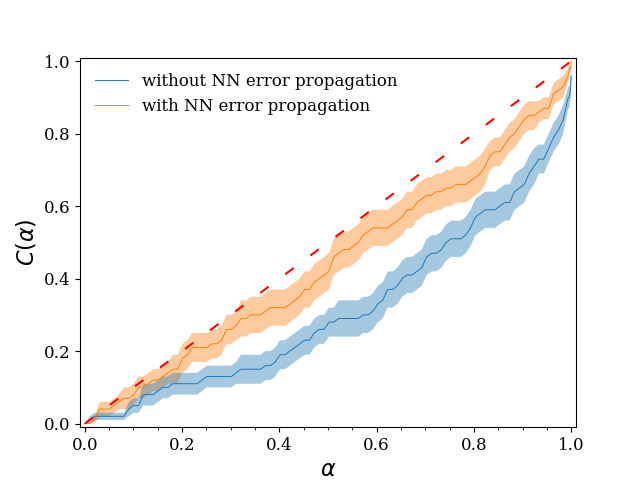}
    \caption{Coverage for inference test from 100 models at $z = 5.4$ uniformly drawn from our priors on $T_0$, $\gamma$, and $\langle F \rangle$ with (orange) and without (blue) NN error propagation in likelihood. Red dash line is where $C(\alpha) = \alpha$, and the shaded region shows the Poisson errors. }
    \label{fig:compare_cov}
\end{figure}

Orange contour in Figure \ref{fig:compare_cov} shows the coverage test result for inference procedure described in Section~\ref{sec:nn err} on 100 mock data at $z = 5.4$. The Ly$\alpha$ forest is generally a non-Gaussian random field, but the autocorrelation function of it averages all pixel pairs over velocity bins which should make the mock draws Gaussian distributed about the mean for each bin. A careful investigation of this assumption has been conducted in appendix C of \citetalias{wolfson2023forecasting}. Hence, as long as our emulator is doing a statistically good job at predicting Ly$\alpha$ autocorrelation function, the inference test should reflect a small deviation from $C(\alpha) = \alpha$ that follows a similar shape to that in \citetalias{wolfson2023forecasting}, resulting from the non-Gaussian distribution of mock draws at high $z$. Appendix~\ref{appdix:cov_plots} shows coverage plots at other higher redshifts, demonstrating that the inference test has consistent performance on mocks from the same seed of random draws from the parameter priors at each $z$.

The non-Gaussian distribution of \lya~autocorrelation function for thermal models at high $z$ is discussed in full details in \citetalias{wolfson2023forecasting}. Generally speaking, the greater deviation
from a multi-variate Gaussian distribution at higher $z$ exists because of low $\langle F \rangle$. However, our inference process still matches with the NGP model performance while discarding the re-weighting post-process which would introduce an additional source of uncertainty to the posterior distribution \citepalias{wolfson2023forecasting}. In order to compensate for this deviation, we run inference tests again on the 100 models with the same thermal parameters but generate random mocks from a multi-variate Gaussian distribution with the given mean model and covariance matrix. We pass the inference test with these Gaussianized mocks, as demonstrated in Appendix~\ref{appdix: gaussian_data}.

To examine the effect of the NN error propagation, an inference test is conducted on the same 100 sets of mock data without the error correction in likelihood, shown as the blue coverage contour in Figure~\ref{fig:compare_cov}. A large discrepancy between the two coverage shows that propagation of NN emulator error is necessary even with an accuracy of $\sim0.5\%$. Taking into account the prediction error, the inference results are more statistically correct. We therefore conclude that the NN emulator gives reliable posterior contours that reflect the underlying thermal information of any mock \lya~autocorrelation function with imperfect resolution and flux levels as described in Section \ref{subsec:sims}.

\subsection{Thermal Evolution Measurement}
For a noisy mock data set with its covariance, we can now perform the HMC inference as described in Section~\ref{sec:inference} and recover its thermal model. To test whether the NN emulator accurately infers the thermal evolution of the IGM in 7 individual redshift bins with $5.4 \le z \le 6.0$, measurements of $T_0$, $\gamma$, and $\langle F \rangle$ in each $z$ are obtained by applying the inference procedure to the \lya~autocorrelation data with thermal parameters at the centers of each parameter grid (as listed in Tabel~\ref{tab:central vals}).

\begin{figure}
	\includegraphics[width=\columnwidth]{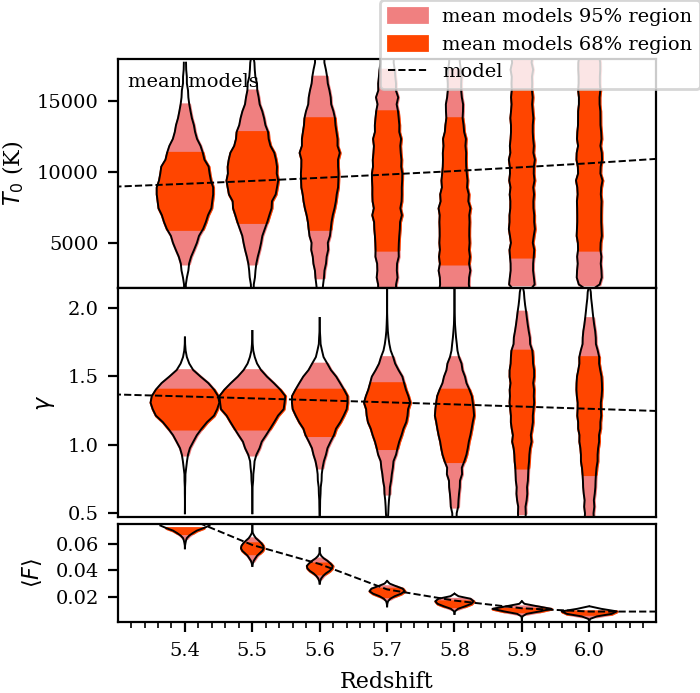}
    \caption{
        The marginalized posteriors for mean model at each $z$ for $T_0$, $\gamma$, and $\langle F \rangle$ as in Table~\ref{tab:central vals}. For each posterior, the light red shaded region demarcates the 2.5th and 97.5th percentile (2$\sigma$) of the HMC draws while the darker red shaded region demarcates the 17th and 83rd percentile (1$\sigma$) of the HMC draws. The true parameter values of the mock data varies as shown by the black dashed lines.
        }
    \label{tab:measurements}
\end{figure}

Figure $\ref{tab:measurements}$ reports the measurements that result from the mean model of the autocorrelation function at each $z$ as an ideal situation, which removes the uncertainties in choosing a random mock and gives the optimal precision of the marginalized posteriors. Translated into the constraints of 68\%, our priors have $\Delta T_0 = \SI{5984}{\kelvin}$, $\Delta \gamma = 0.598$ and $\overline{\Delta \langle F \rangle} = 0.0095$ across redshift \citep{gaikwad_2020, bosman_2021_data}, so our measurements are self-consistent in the constraints. All inferred posteriors contain the true values of $[T_0, \gamma, \langle F \rangle]$ within their $1\sigma$ error bars for these true models. At $z = 5.4$, the posterior measurement constrains $\langle F \rangle$ to $5\%$, $T_0$ to $22\%$, and $\gamma$ to $7\%$ averaging across the 100 mocks from the aforementioned inference test. These constraints demonstrate a constraining power comparable to \citetalias{wolfson2023forecasting}. In general, uncertainties grow monotonically with $z$ which  results from low $\langle F \rangle$.

\begin{figure}
	\includegraphics[width=\columnwidth]{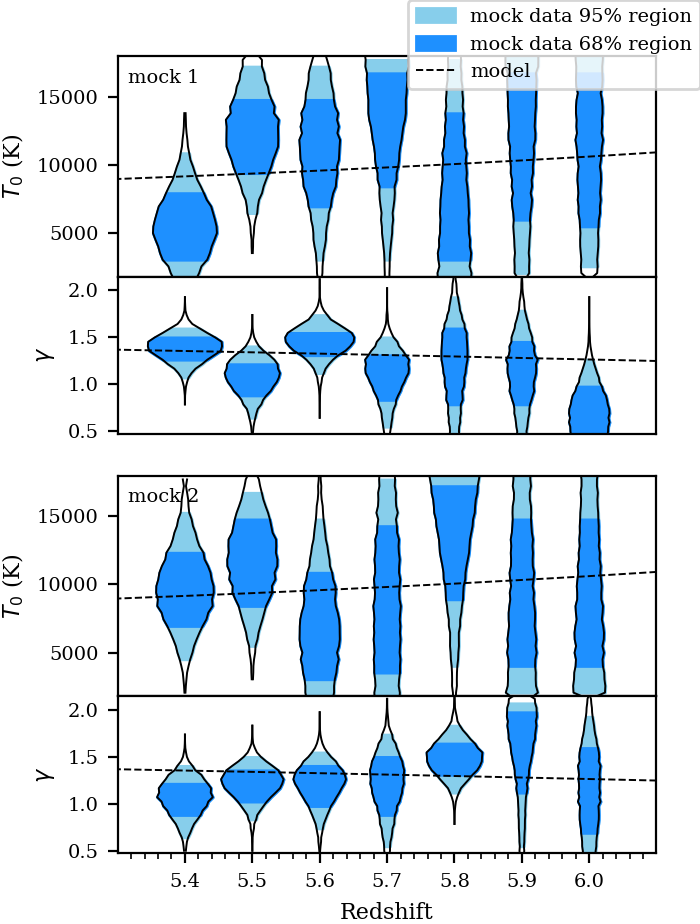}
    \caption{
        The marginalized posteriors for two random mock data sets at each $z$ for $T_0$ and $\gamma$ as in \citetalias{wolfson2023forecasting}. 
        The first and third panels show the marginalized posteriors for $T_0$ while the second and fourth panels show the same for $\gamma$. 
        For each posterior, the light blue shaded region demarcates the 2.5th and 97.5th percentile (2$\sigma$) of the HMC draws while the darker blue shaded region demarcates the 17th and 83rd percentile (1$\sigma$) of the HMC draws. 
        There are 14 total random mock data sets used to make this figure as for each of the 7 redshifts there are 2 random mocks.
        The shape of each posterior is partially determined by the luck of the draw when selecting the mocks. 
        The true parameter values of the mock data varies as shown by the black dashed lines
        as reported in Table \ref{tab:central vals}.    
        }
    \label{fig:temp_gamma_violin_plot}
\end{figure}

Figure~\ref{fig:temp_gamma_violin_plot} visualizes the above results with two random mock data sets at each redshift. The exact two mocks are drawn in figure 9 of \citetalias{wolfson2023forecasting}  for direct comparison. It suggests that the NN emulator exhibits a comparable performance to the traditional NGP with MCMC inference method \citepalias{wolfson2023forecasting} while using 10\% of the total simulations. The cost per effective sample of our HMC inference is compared to that of the MCMC from \texttt{EMCEE} \citep{forman_mackey_2013} package and found to be 20 times smaller. Our NN emulator has consistently well performance in the highest redshifts, $z>5.7$, where the true values are still in the $1\sigma$ errors of measurements despite decreasing constraining power.

\section{Conclusions and Discussions} \label{sec:conclusion}
In this paper, we have presented a computationally efficient solution for the IGM thermal parameter inference using realistic mock high-$z$ quasar sightlines. At $z = 5.4$, the NN emulator has a high prediction accuracy for the \lya~autocorrelation function: 99\% of the test set has an emulation error within 2\%, and an average error over the set is $0.56\%$. This emulator error is as low as $\sim3\%$ of the total error budget, sufficiently small for our forward-modelled simulations, designed to mimic realistic high-resolution observational data from echelle spectrographs \citepalias{wolfson2023forecasting}. Training on only 100 models, $10\%$ of the original simulations, establishes the computational advantage of our NN emulator framework. Examining the loss shape, we observe that the smaller velocity bins generally exhibit higher error. This is attributed to the smaller bin size and the more pronounced imprints of the IGM's thermal history in small-scale regime, which makes it inherently harder to predict.

Following the method in Section~\ref{sec:inference}, the thermal parameter measurements exhibit similar performance to \citetalias{wolfson2023forecasting} in terms of the inference test with random mocks of simulation skewers, demonstrating a statistical success in emulating the autocorrelation function from the same simulation models. This method is made more robust by taking into account the uncertainties of the emulator in reconstructing the \lya~autocorrelation functions at a given redshift. The propagation of the emulation error has been proven necessary for an average accuracy as low as $\sim0.5\%$ ($z =5.4$) because for small velocity lags the emulator error reaches $\sim2\%$, while the statistical uncertainties of the data shrink to $\sim10\%$ in this regime. This impacts the extraction of thermal information, which is most pronounced in the smaller velocity regions. For the inference test at different $z$, the suboptimal coverage plot deviations come from the incorrect assumption of the multi-variate Gaussian distributed mock data set, as further demonstrated in Appendix~\ref{appdix: inf_test}. Without this caveat in the likelihood assumption, we successfully pass the inference test, demonstrating that the posteriors provided by the emulator are reliable.

We also compare this NN emulator implementation with the conventional NGP method in terms of time expense and inference accuracy. The total data sets for training and uncertainty approximation of the emulator save approximately 17M GPU hours from additional simulation runs. The overall inference time is two orders of magnitude faster with the differentiable emulator than with the NGP method, while the cost per effective sample is 20 times smaller for HMC compared to traditional MCMC. The constraining power of parameter inference in Table~\ref{tab:central vals} is at the same level as \citetalias{wolfson2023forecasting}, as the precision of the constraints decreases with increasing $z$. These thermal parameter measurements and the coverage test results of the NN emulator demonstrate that it achieves statistically robust performance for inference from mock sightlines with realistic noise.

Future improvement on the emulator includes reducing the systematic emulator uncertainty at small scales by potentially applying non-linear transformations of training data across the velocity range. To obtain precise constraints on the thermal state of the IGM from real-life observational data, incorporating an additional emulator to estimate the covariance during the interpolation would improve the accuracy of uncertainty quantification for each data set. Future work on more realistic UVB models at this redshift range will also be necessary to get the best possible constraints on reionization from this framework, i.e. not only the thermal state of the IGM but also the mean free path of ionizing photons that describes the UVB. A larger than existent number of high-SNR high-resolution QSO surveys at this high-$z$ range will be required to implement our NN emulator for making real measurements of the thermal history and sampling the statistical uncertainties related to the autocorrelation function. We expect such realistic observational data to be obtained from future echelle spectrographs, e.g. Keck/HIRES, VLT/UVES, VLT/XSHOOTER, and Magellan/MIKE. 

\section*{Acknowledgements}
We acknowledge insightful discussions with the ENIGMA group at UC Santa Barbara and Leiden University. JFH acknowledges support from the National Science Foundation under Grant No. 1816006.

This research used resources of the National Energy Research Scientific Computing Center (NERSC), a U.S. Department of Energy Office of Science User Facility located at Lawrence Berkeley National Laboratory, operated under Contract No. DE-AC02-05CH11231.

\section*{Data Availability}
The code and simulation data presented in this paper are available for reproducing results upon reasonable request.

%contained in the following GitHub repository by clicking the icon: \href{https://github.com/enigma-igm/igm_emulator.git}{\faGithubSquare}.



\bibliographystyle{mnras}
\bibliography{main} % if your bibtex file is called example.bib

%\begin{thebibliography}{99}
%\end{thebibliography}

%%%%%%%%%%%%%%%%%%%%%%%%%%%%%%%%%%%%%%%%%%%%%%%%%%

%%%%%%%%%%%%%%%%% APPENDICES %%%%%%%%%%%%%%%%%%%%%

\appendix
\section{Uniform UVB and Homogeneous Reionization}\label{appdx:UVB}

Recent observations indicate that the UVB cannot be well described by uniform fields for $z \gtrsim 5.0$ \citep{becker_2021, bosman_2021_limit, gaikwad_2023, zhu_2023}. In this work, we used simulations based on a uniform UVB and instantaneous reionization model for $5.4 \leq z \leq 6.0$  because this redshift range represents the lowest available simulations analyzed in \citetalias{wolfson2023forecasting}, providing a baseline for directly comparing this framework to previous performance while requiring fewer simulations.

\citet{onorbe_2019} showed that UVB fluctuations and temperature fluctuations manifest on large scales ($k \sim \SI{1e-3}{\second\per\kilo\meter}$) while the small-scale ($k> \SI{0.06}{\second\per\kilo\meter}$) correlation properties of the \lya~forest are not significantly affected. Since the thermal state of the IGM described in Equation~\ref{eq:temperature-density relation} sets the small-scale power, the differences of assuming a uniform UVB are not fully captured in the scale of interest. To further demonstrate this point, the effect of temperature and UVB fluctuations on the \lya~forest flux autocorrelation function has been explored in the last section of \citetalias{wolfson2023forecasting}, which analytically shows that adding UVB fluctuations would add a slight boost at large scales of the correlation function, though investigating this further is beyond the scope of the paper. However, future studies can apply this framework to more realistic reionization and UVB models to describe the IGM at $z \gtrsim 5.0$.

\section{Emulator details}

\subsection{Splitting data for training, test, and validation}\label{sampling}
To make our grid of thermal models, we use 15 values of $T_0$ and 9 values of $\gamma$ resulting in 135 different combinations of these parameters at each $z$ \citepalias{wolfson2023forecasting}. The mean transmitted flux, $\langle F \rangle$, has 9 values to vary for each model, so in total we have 1215 grid points of simulations. Because this grid of parameters is not evenly spaced, it's not ideal to sample with Latin Hypercube in interval scaling methods. Instead, a regular mesh grid and a random split function from \texttt{Tensorflow} are used to tackle the problem by directly sampling a reduced-dimensional regular mesh grid through interpolation of the nearest grid point. We apportion the data into different sets as follows: 50\% of the total data are sampled for the training set, 40\% for the validation set, and the rest 10\% for the test set. All data used for training the NN emulators are pre-processed by standardization, i.e., dividing
each thermal parameter or correlation function by its standard deviation after subtracting its mean. This ensures a more rapid convergence of training \citep{Wan_2019}.

\begin{figure*}
     \centering
     \begin{subfigure}[h!]{\textwidth}
         \includegraphics[width=0.25\textwidth]{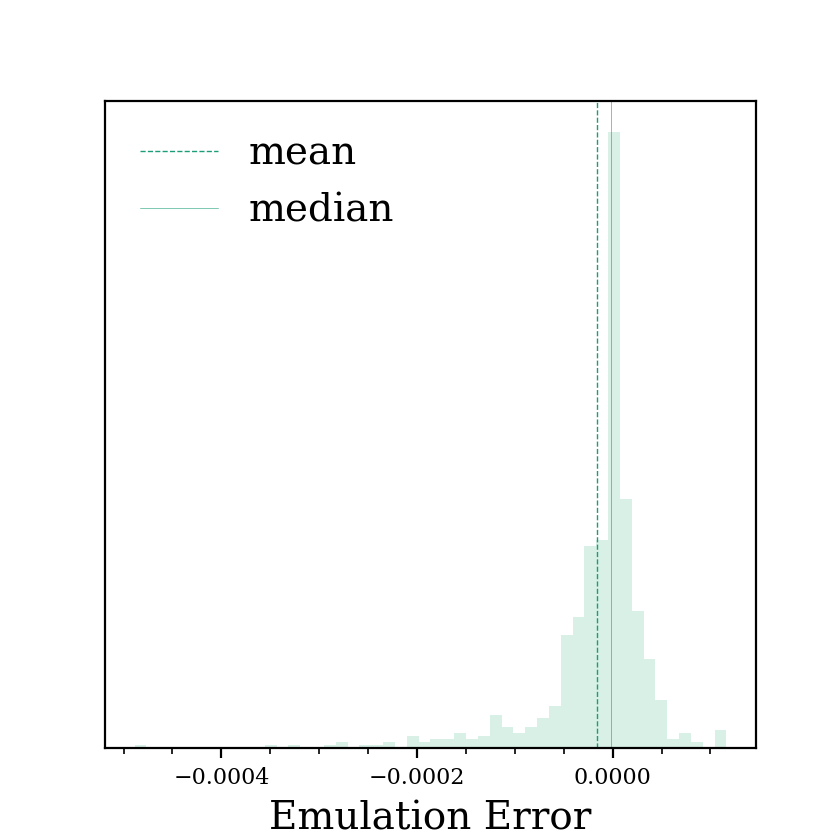}
          \includegraphics[width=0.25\textwidth]{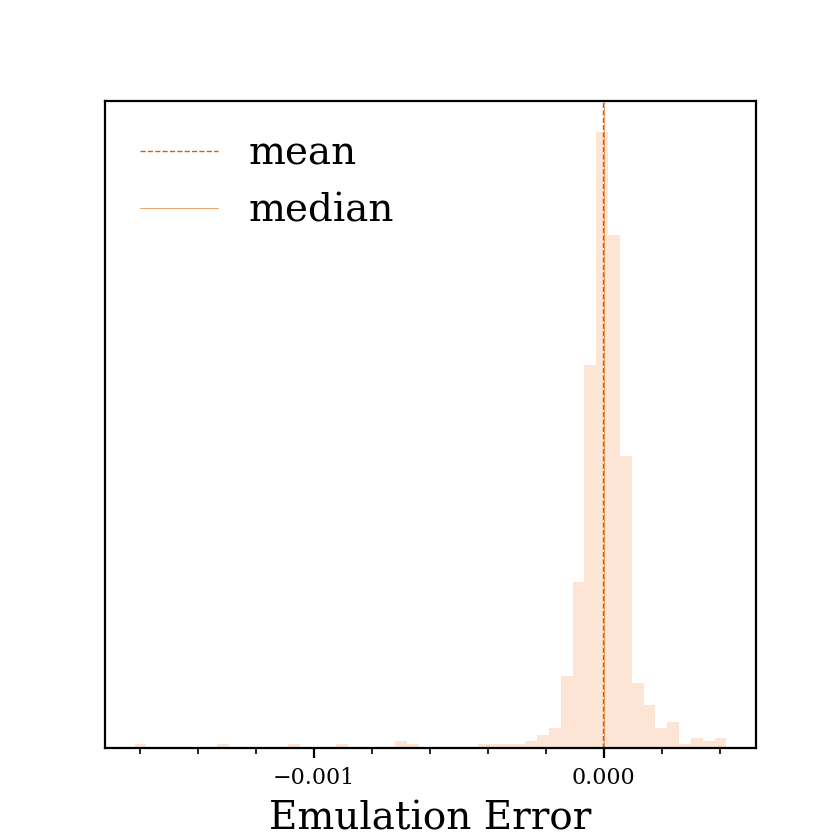}
         \includegraphics[width=0.25\textwidth]{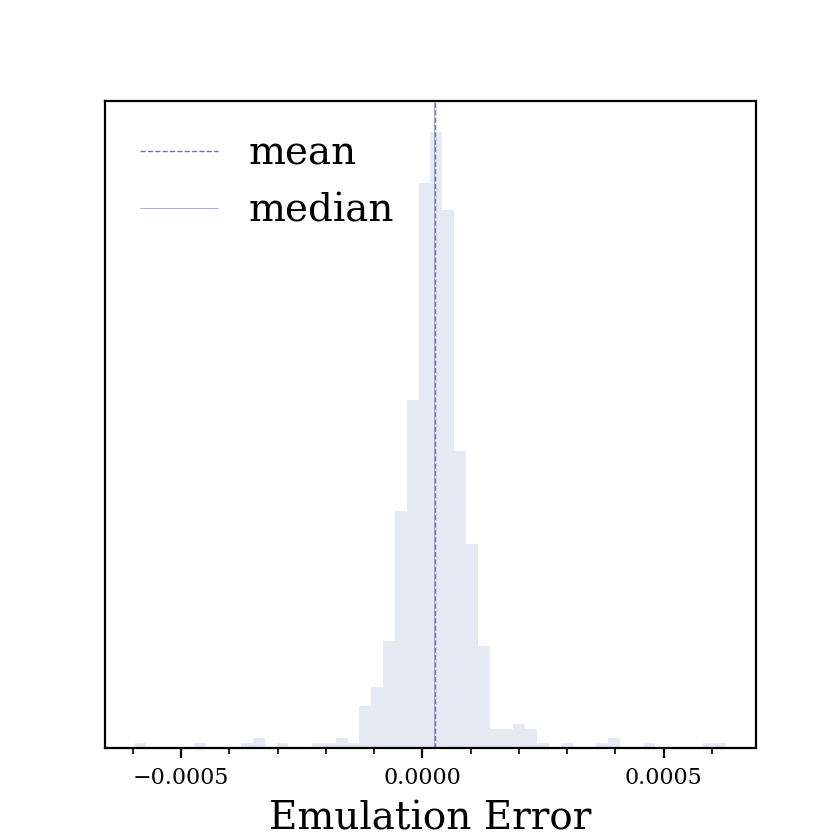}
         \includegraphics[width=0.25\textwidth]{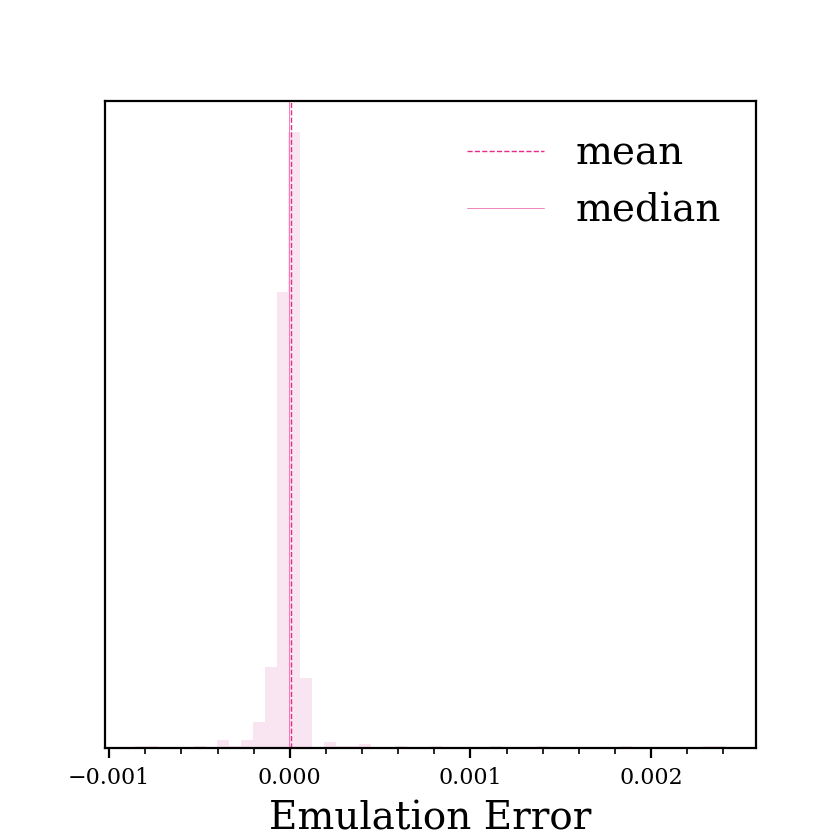}
         \caption{12 autocorrelation functions for the test set with different sampling seeds in random split from left to right.}
     \end{subfigure}
     \hfill
     \begin{subfigure}[b]{\textwidth}
        \includegraphics[width=0.25\textwidth]{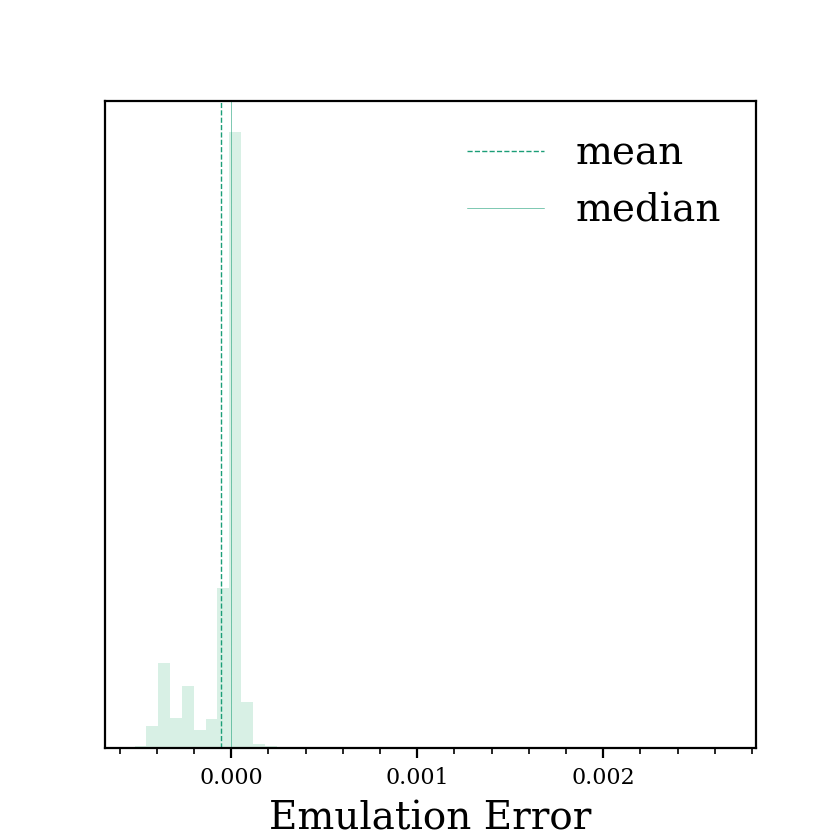}
          \includegraphics[width=0.25\textwidth]{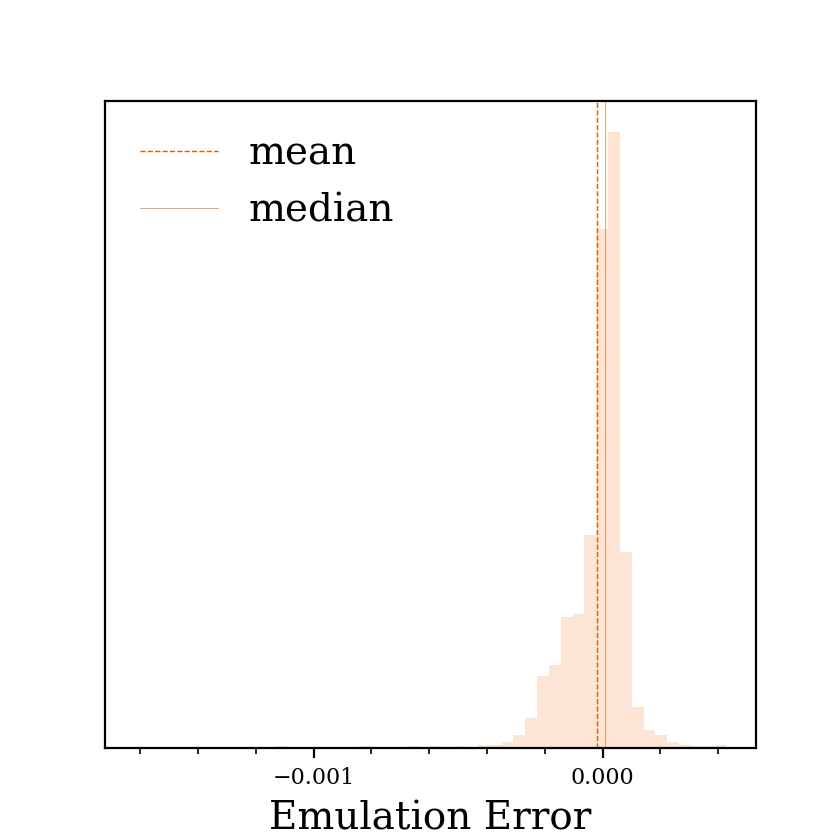}
         \includegraphics[width=0.25\textwidth]{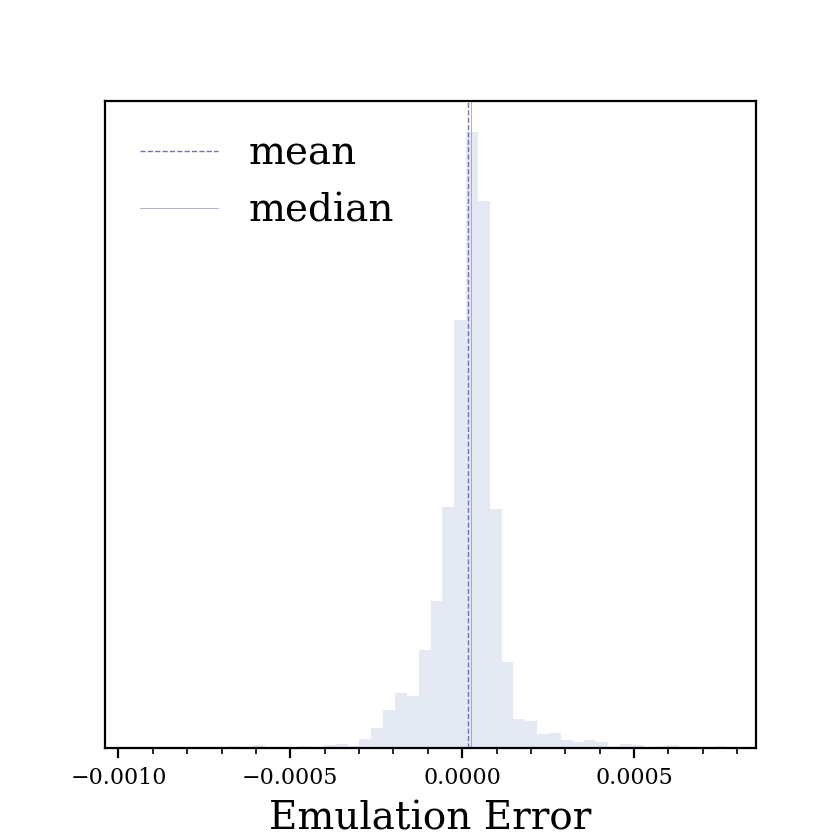}
         \includegraphics[width=0.25\textwidth]{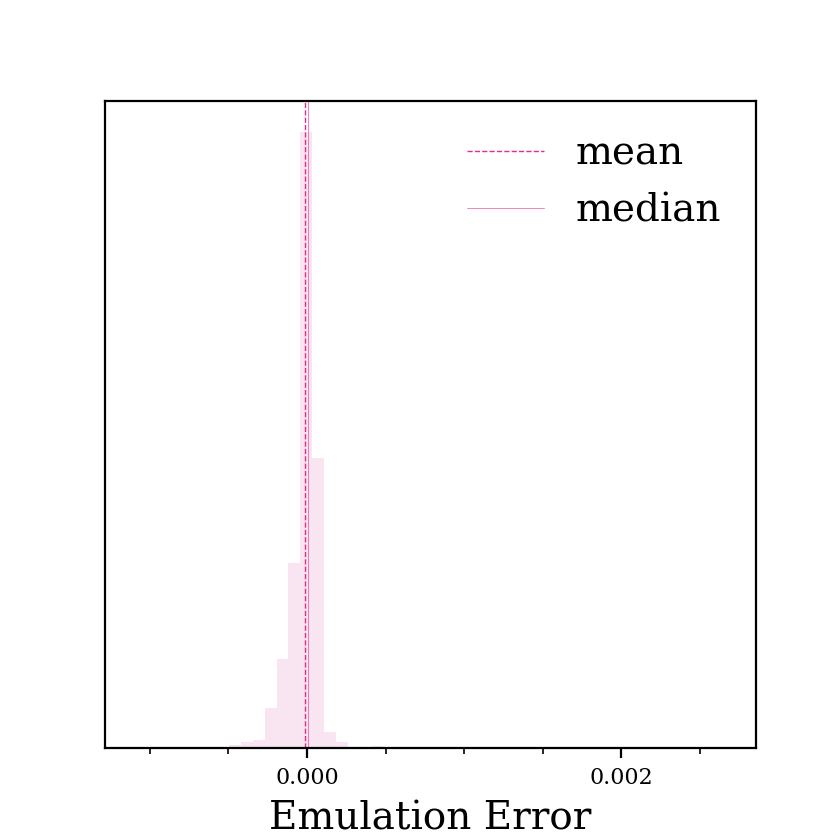}
         \caption{62 autocorrelation functions for the test set with different sampling seeds in random split from left to right.}
     \end{subfigure}
        \caption{Gaussianity test for the error approximation for different sizes and sampling seeds, where same color of histogram denotes same seed for sampling.}
        \label{fig:gaussian_err}
\end{figure*}

An additional set of test data can be sampled separately to add to the test set for error approximation in Section~\ref{sec:nn err}, while the metric for emulator performance is still evaluated with the original test set for consistency. To test the Gaussianity of $\mathbf{\Delta_\text{NN}}$, we sampled different sizes of the test set with different seeds. Figure~\ref{fig:gaussian_err} shows the histograms of $\mathbf{\Delta_\text{NN}}$ at $z = 5.4$, as a very basic test, where the mean and median are  expected to overlap for a normally distributed sample \citep{Grand_n_2022}. The size of test data set turns out to make trivial difference, so we use the smallest size. The total data set size is 112, comprising 12 autocorrelation functions for the test set, 55 for the training set, and 45 for the validation set, as shown in Figure~\ref{fig:train data}. This data set supports the emulator's performance evaluation across all redshifts.

\begin{figure}
 \includegraphics[width=\columnwidth]{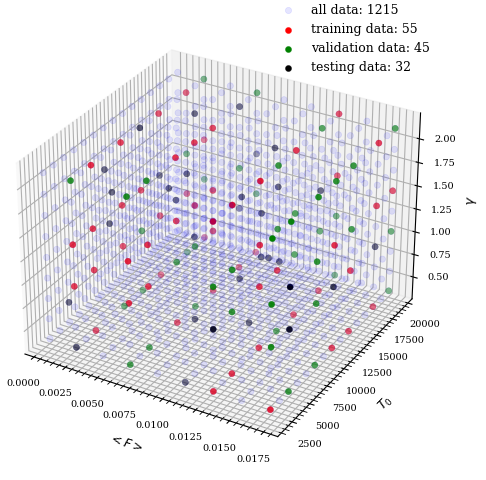}
  \caption{Data split in thermal parameter grid of $[T_0, \gamma, \langle F \rangle]$ for $z = 5.9 - 6.0$. Both the training and validation sets are kept at same positions in parameter space as in Figure~\ref{fig:train data} for each redshift $z = 5.4 - 6.0$, whereas the $z = 5.9 - 6.0$ the test data set have additional 20 points for error estimation due to the larger uncertainties in emulator predictions. }
\label{fig:train data high z}
\end{figure}

Note that for high redshift, $z = 5.9, 6.0$, the first row of mean transmission flux is excluded from the training and validation sets for the reason that the increasingly small mean flux ($\sim10^{-4}$) would result in extremely non-linear noise at small velocity bins and affect the training stability. As a result, the prediction error from the emulator is higher, and thus requires more test data for the error approximation. As in Figure~\ref{fig:train data high z}, the test data size for high $z = 5.9 - 6.0$ is 32. Despite the increase in number of models needed for error approximation, the emulator is still computationally efficient when compared to constructing a fine grid of models. 

\subsection{Hyperparametor choices}
\label{hparam}
As explained in Section~\ref{mlp}, there are a few hyperparameters in the MLP architecture. The ones we tuned on with \texttt{Optuna} are layer sizes, learning rate settings, and batch size.

\textbf{Loss function:} Various loss functions were experimented with, i.e., Mean Squared Error (MSE) (with and without Fourier transformation error in discrete cosine functions), Mean Chi (using a fixed covariance matrix at the center of parameter grid), Huber Loss, Mean Absolute Error (MAE), and Relative
Mean Absolute Error (RMAE). We looked into different loss functions in order to figure out a best way to handle the uneven scale of the data across velocity bins and also regularize learning towards small weights to prevent overfitting to complex structure with L2 regularization. In order to take account of the scale of the physical data, the loss was calculated in physical units instead of the standardized ones passed through the training loop. In the end, we chose the RMAE loss function because it's faster to compute compared to taking inverse of a covariance matrix in both Mean Chi loss and Huber Loss, and it also weighs the loss on different velocity bins with percentage error compared to MSE and MAE. The L2 term is removed from the original loss function as it does not behave as intended for adaptive gradient algorithms such as Adamw we use \citep{loshchilov2019decoupled}.

\textbf{Learning rate:} We optimise the Mean Absolute Percentage Error loss function using Adamw \citep{deepmind2020jax} which uses weight decay to regularize learning towards small weights. Note that this weight decay is multiplied with the learning rate. Apart from its decay parameter, the learning rate was input with a descending function \texttt{optax.warmup\_cosine\_decay\_schedule} that has a linear warmup followed by cosine decay. Additionally, to further avoid  training instabilities, a gradient clipping mechanism was designed to constrain the maximum gradient leap \citep{clipping}. With \texttt{Optuna}, the optimal setting for maximum gradient norm is $0.4$ and for weight decay regularization is $0.003$. For $z = 5.4$ emulator, the initial learning rate is $0.005$.

\textbf{Layers:} The input layer has dimension of $3$ while the output layer has $59$ for velocity bins of \lya~autocorrelation function. We use 2 hidden layers with 100 perceptrons each after hyperparamter tuning.

\textbf{Activation function:} The activation functions we explore include \texttt{'jax.nn.leaky\_relu'}, \texttt{'jax.nn.relu'}, \texttt{'jax.nn.sigmoid'}, and \texttt{'jax.nn.tanh'}, in which \texttt{Optuna} optimization chooses 
\texttt{'jax.nn.tanh'} in Equation \eqref{eq: tanh}, which is the Hardtanh, a cheaper and more computational efficient version of tanh. The main advantage provided by the function is that it produces zero centred output thereby aiding the back-propagation process \citep{nwankpa2018activation}.

\begin{equation} \label{eq: tanh}
    \mathrm{hard\_tanh}(x) = \Biggl\{
    \begin{aligned}
        -1, ~&x < -1 \\
        x, ~&-1 < x < 1 \\
        1, ~&x > 1
    \end{aligned} 
\end{equation}

\textbf{Training epochs and early stop:} The emulator is trained for 2,000 epochs within 5 seconds. To prevent over-fitting during this many epochs, an early stop mechanism stops the training when the validation loss is not improving for more than 200 epochs. For $z>5.8$, the early prevention value is set to 500 instead so the training epochs last longer for the increasing complex shape of the autocorrelation function.

\textbf{Batch size:} For each epoch, the emulator is trained on random mini batches of the training data set to accelerate the process. For a total of 112 models, we test on sizes [None, 32, 50] (None denotes training on the entire training set) and marginalize on the validation loss with a batch size 50. The purpose of mini-batching the training data is to eliminate the need for Dropout, be less careful about weight initialization, and speed up the gradient descent computation \citep{ioffe2015batch}.

\section{HMC inference procedure} \label{appdix:HMC_setting}

The \texttt{numpyro.infer.NUTS} sampler kernel was initialized with with 4 chains of 4,000 samples following 1,000 warm-up samples running in parallel on a single device with the "vectorized" drawing method. The \texttt{max\_tree\_depth} was set to $10$ (i.e., 1024 steps for each iteration). The \texttt{potential\_fn} has input of the sum of error propagated log-likelihood function and log-priors. The \texttt{target\_accept\_prob} is set to 0.9, instead of the default 0.8, for a smaller step size and more robust sampling \citep{hoffman2014no}. All parameters in the inference process have been transformed from a bounded parameter vector $\boldsymbol{\theta}$ into an unbounded parameter vector $\boldsymbol{x}$ using a logit transformation. The purpose of this transformation is to have unbounded priors and thus a continuum and differentiable probability.

For a complete inference procedure with HMC, a Ly$\alpha$ autocorrelation function data set is passed as input $\boldsymbol{\xi}$ into Equation \eqref{eq:nn error}. Under least-square fitting with the emulator, we have an optimal thermal parameter point as the initial position for sampling $\boldsymbol{\theta}_i$, and each point we can emulate the data as $\boldsymbol{\xi_\text{NN}(\boldsymbol{\theta})}$ to calculate the likelihood. \texttt{NUTS} starts to sample and either accepts or rejects proposed parameter points based on the likelihood surface. After 4,000 trials, marginalization on the potential is achieved with inference measurements in the 50th percentile of all the posteriors. 

\section{Performance at other redshifts} \label{appdix:other_z}
\begin{figure*} 
\includegraphics[width=\columnwidth]{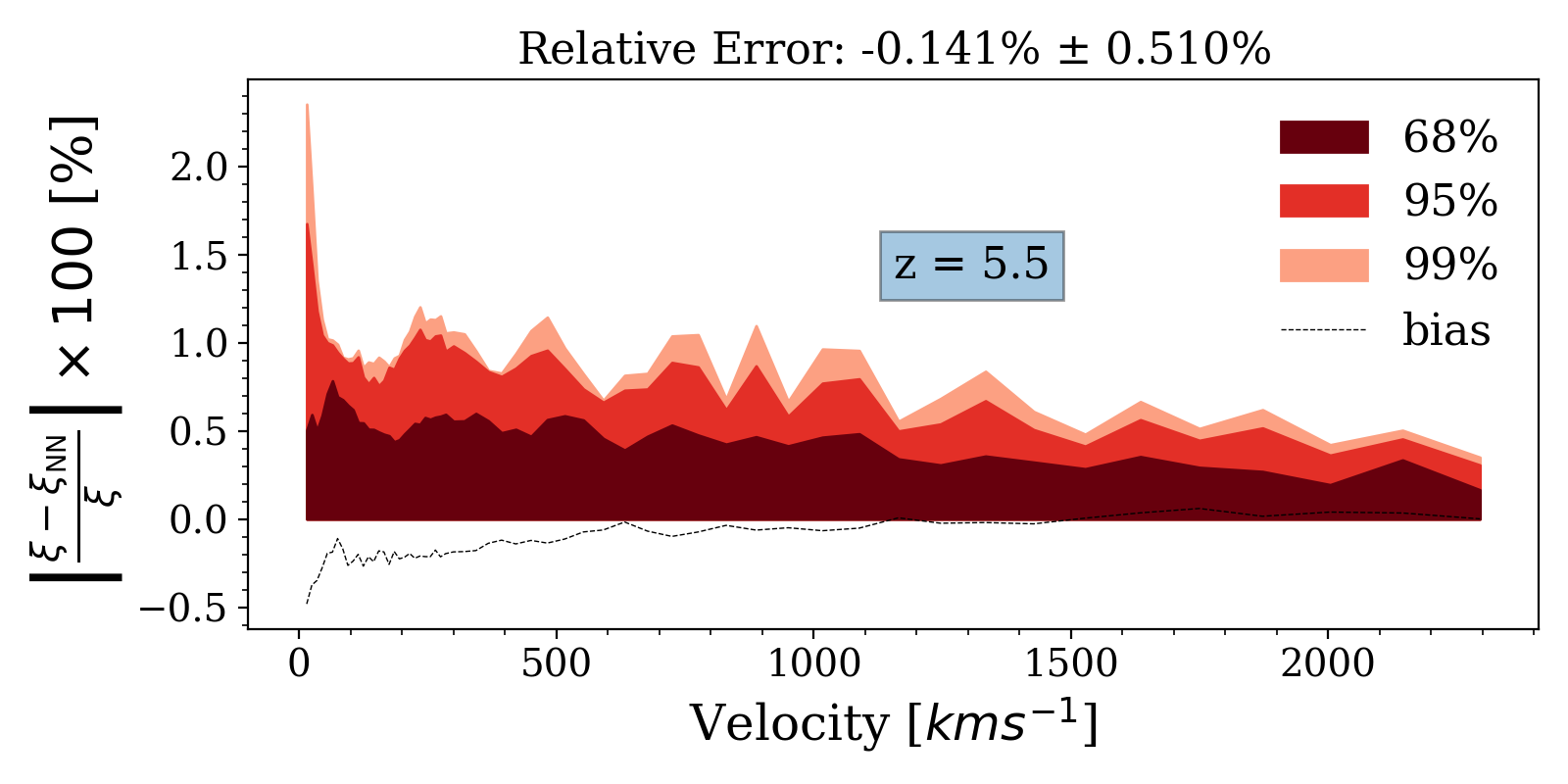}
\includegraphics[width=\columnwidth]{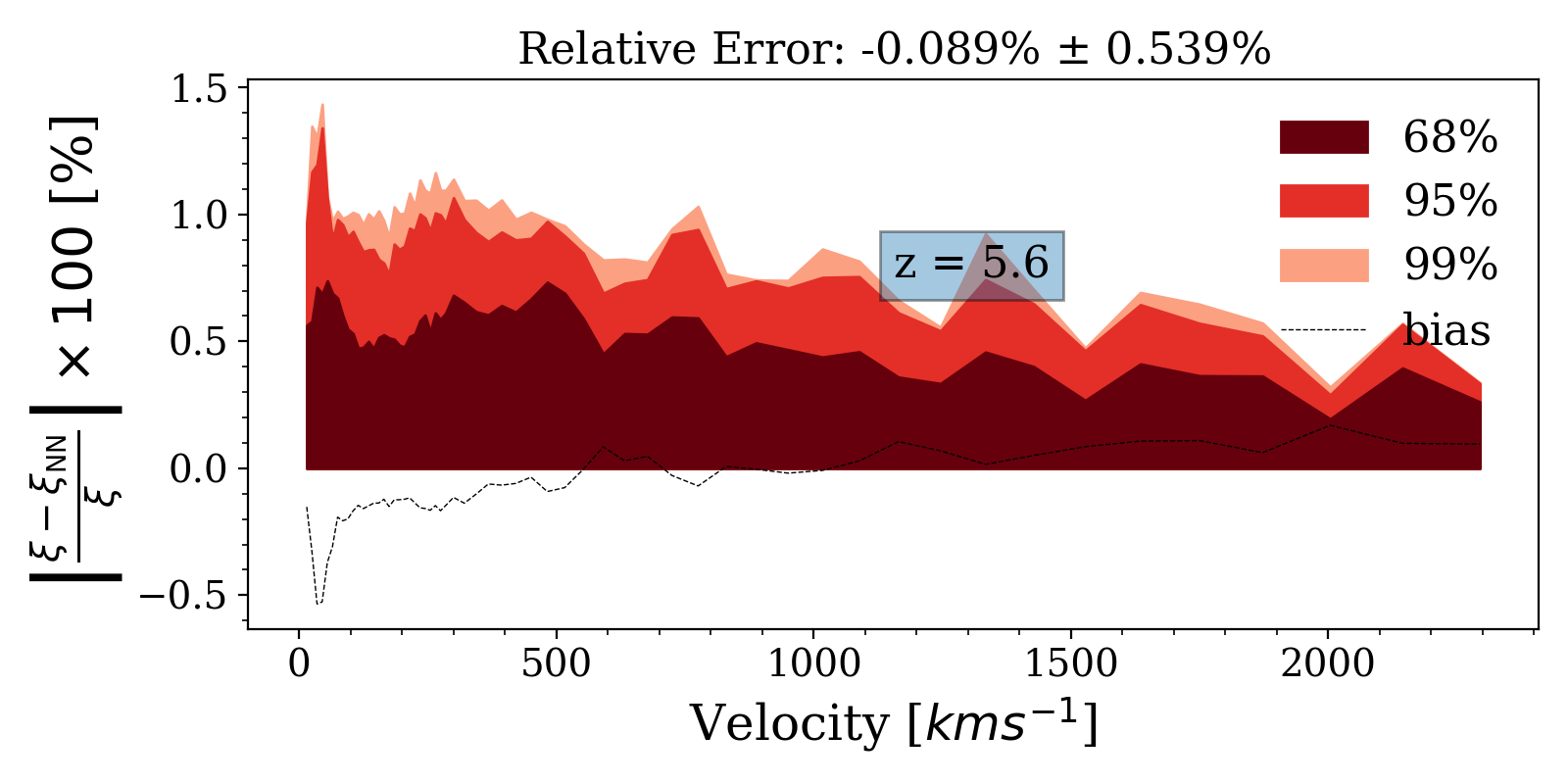}
\includegraphics[width=\columnwidth]{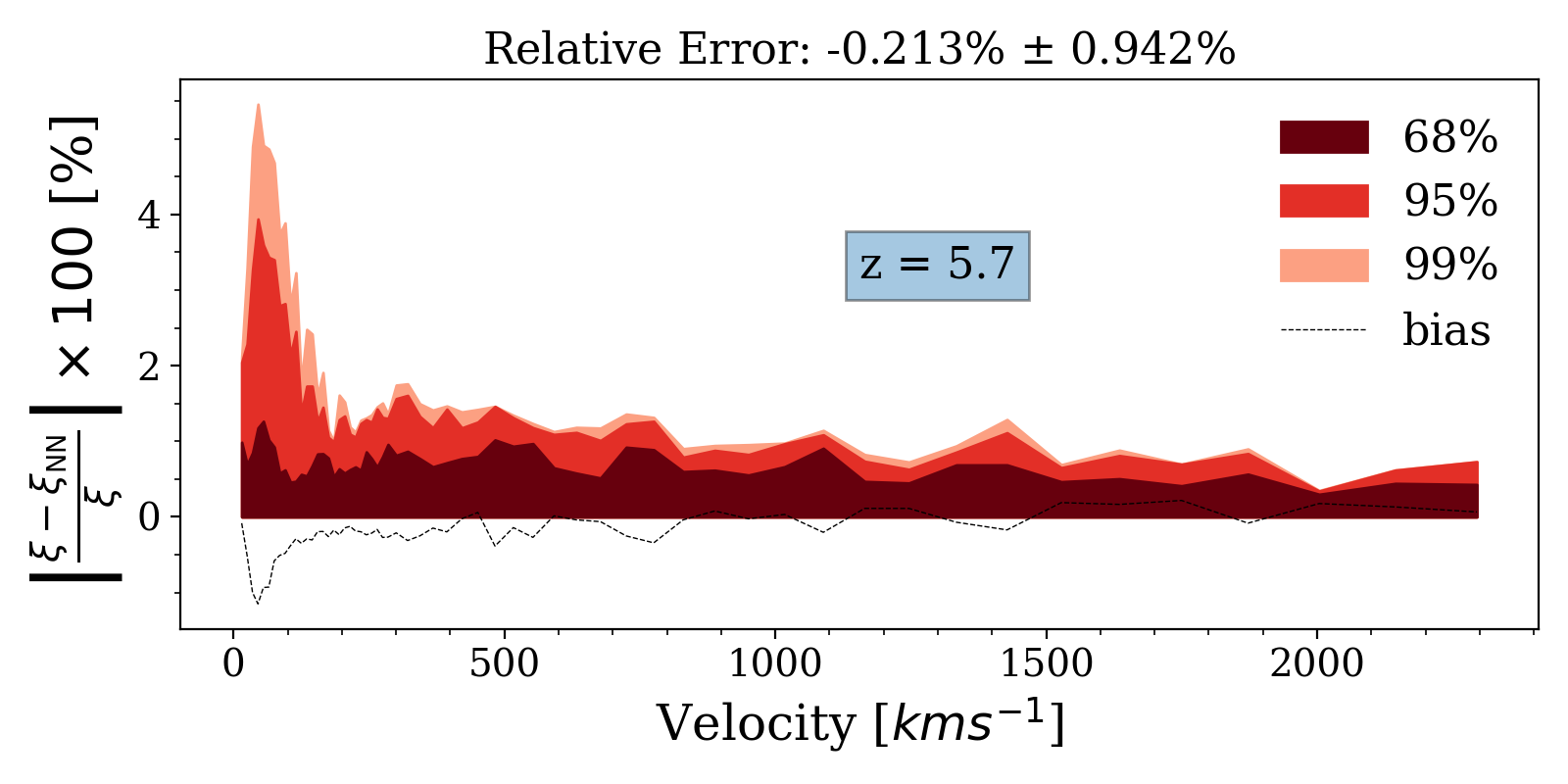}
\includegraphics[width=\columnwidth]{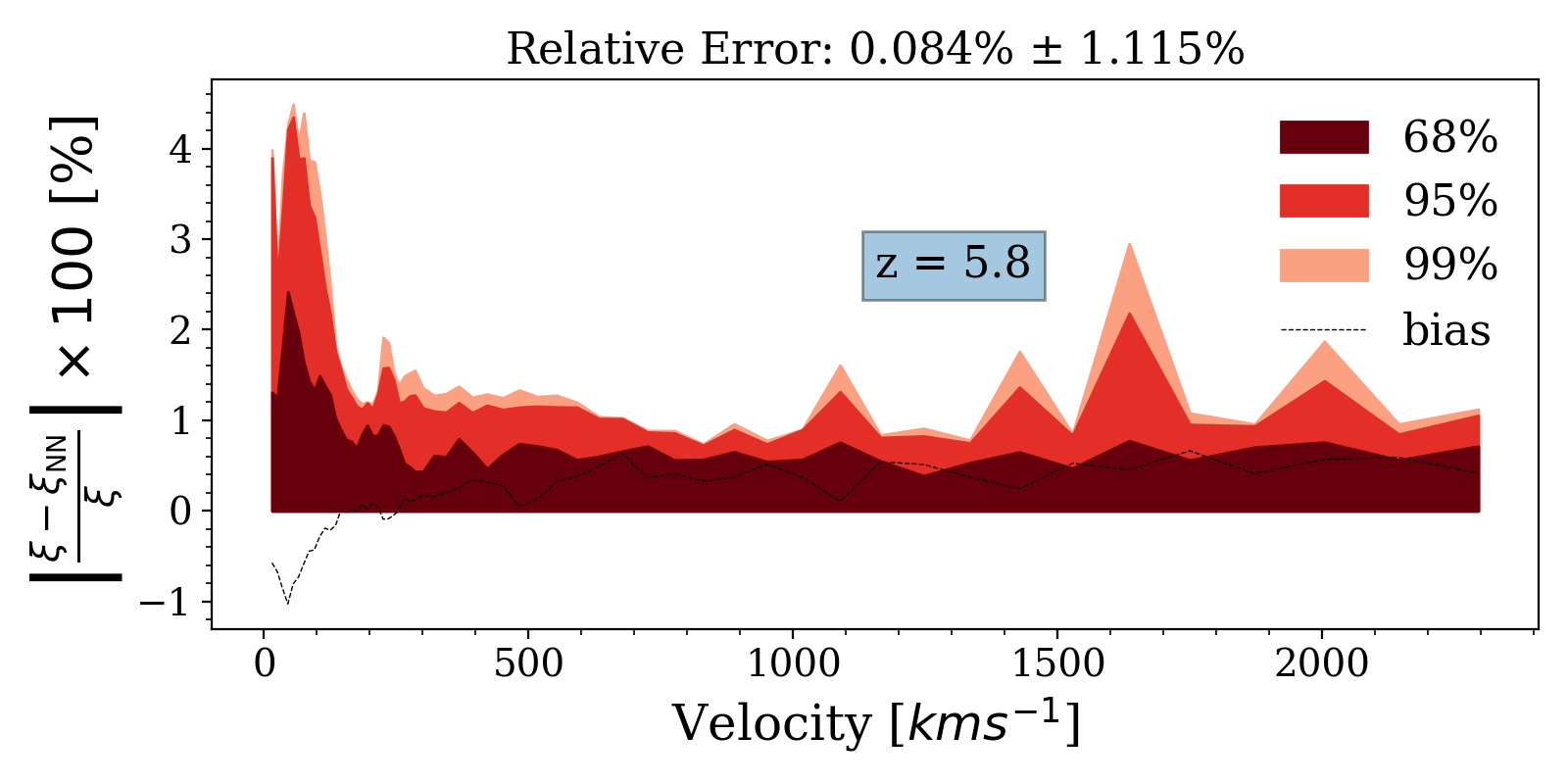}
\includegraphics[width=\columnwidth]{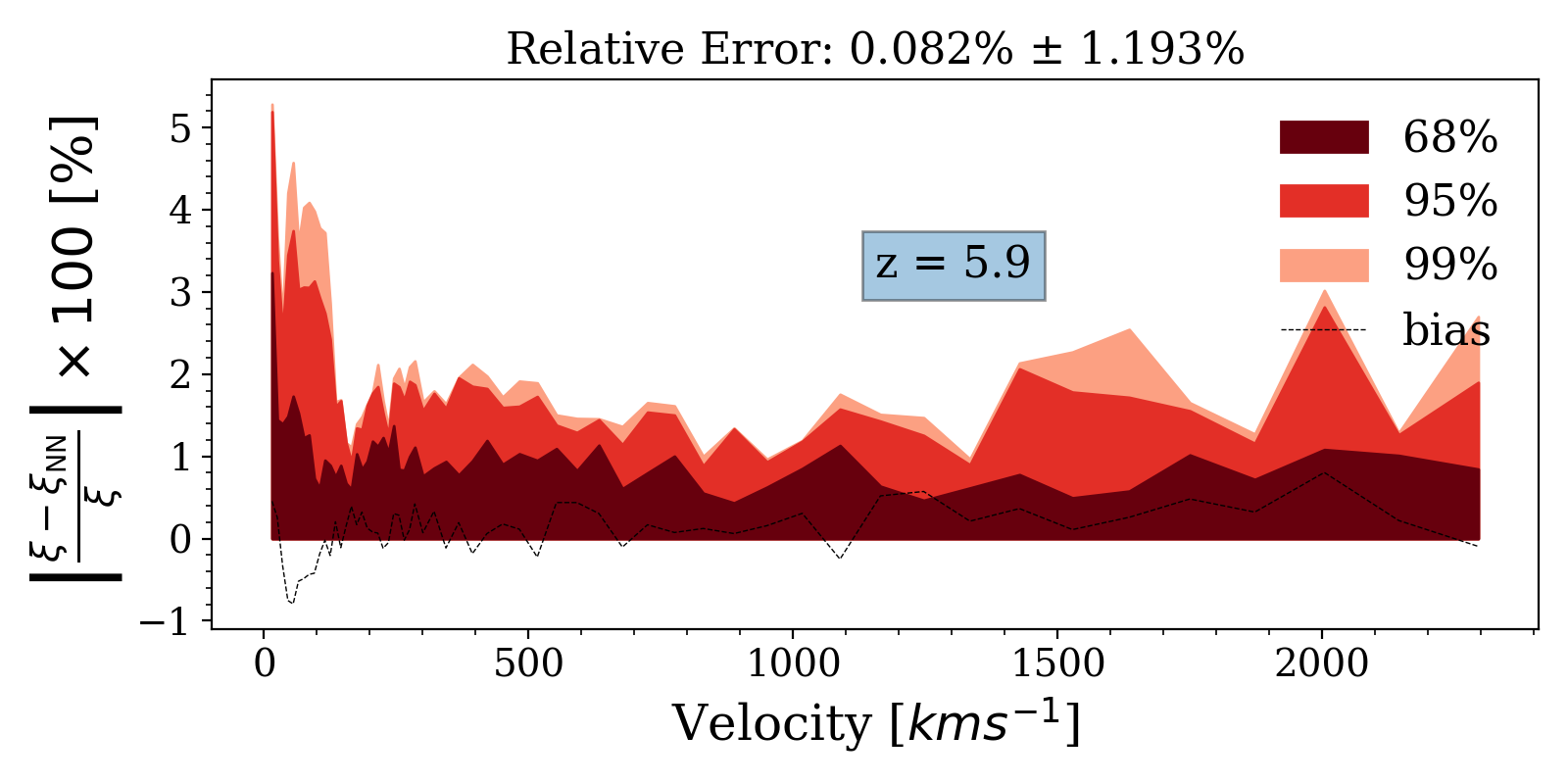}
\includegraphics[width=\columnwidth]{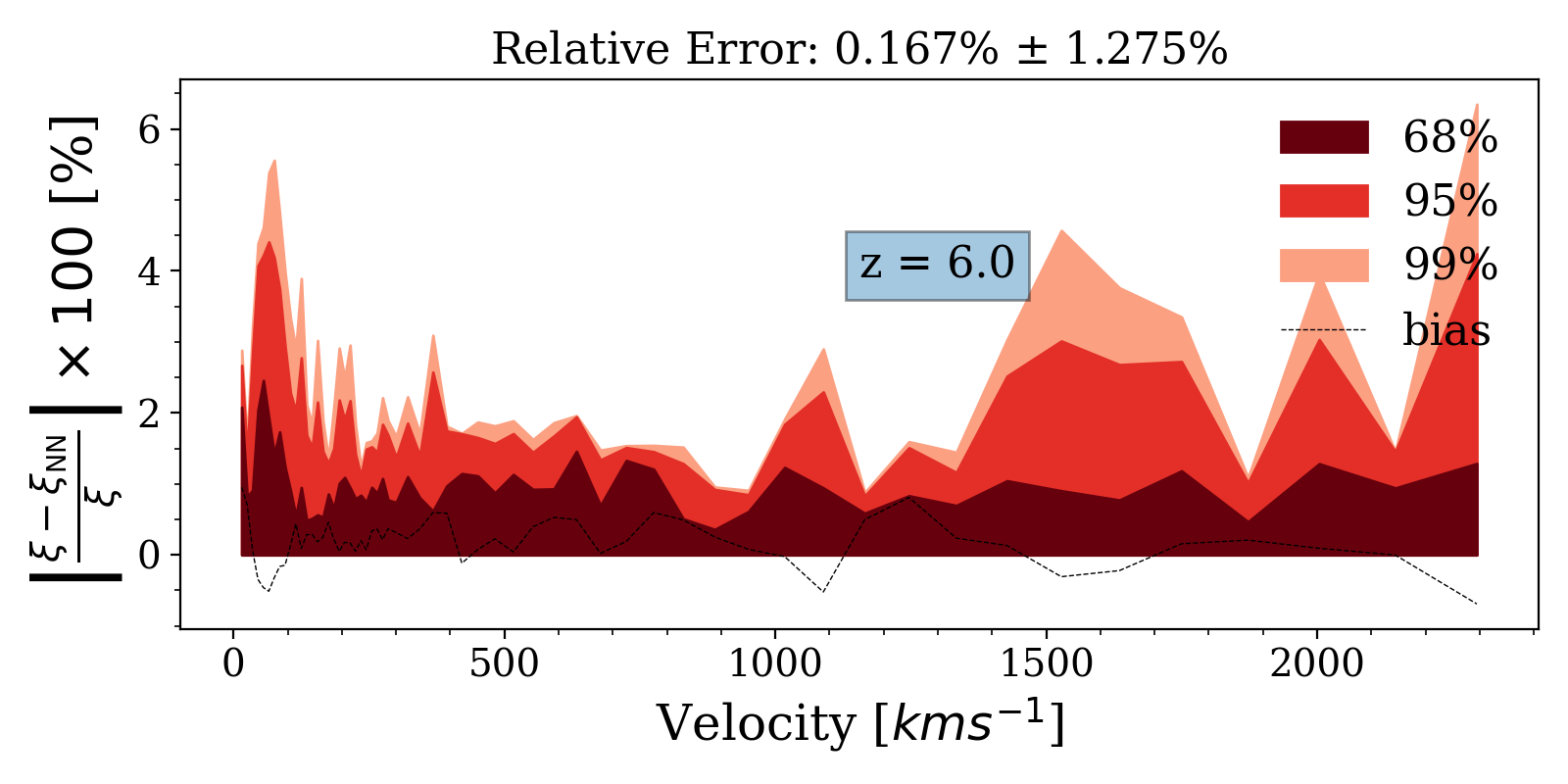}
\caption{Emulation error for $z = 5.5-6.0$, redshift  labeled in blue text boxes. It shows bias (dotted line) and standard deviation ($68\%$ region) of the relative percentage error evaluated from the 12 \lya~test data set. For most redshift, the NN emulator meets the percent-level error while the uncertainties increase with redshift.} \label{fig:all z emu error}
\end{figure*}

This section shows the performance of emulators at other redshift ($z = 5.5-6.0$) with the percentiles of the emulation error for the test set in Figure~\ref{fig:all z emu error}. As discussed in Section~\ref{sec:conclusion}, higher the redshift is, higher the emulation error is, and hence the increasing residual dynamical range in Figure~\ref{fig:all z emu error} is expected. As in Section \ref{sampling}, by excluding the first values of the mean transmission flux in $z = 5.9-6.0$, the emulators converge in a reasonable elapsed time while maintaining a percent level error. But the missing data of $\langle F \rangle_{\mathrm{min}}$ models in training manifest in the poor fitting of the emulation. For performance at $z = 5.9 - 6.0$, models with $\langle F \rangle_{\mathrm{min}} = 0.0006$ at $z = 5.9$ and $\langle F \rangle_{\mathrm{min}} = 0.0005$ at $z = 6.0$ unstabilize the overall prediction accuracy because the emulator tends to overestimate the mean flux.

\section{Inference test} \label{appdix: inf_test}
\subsection{Gaussian Data Inference test}\label{appdix: gaussian_data}
\begin{figure*}
   \includegraphics[width=\columnwidth]{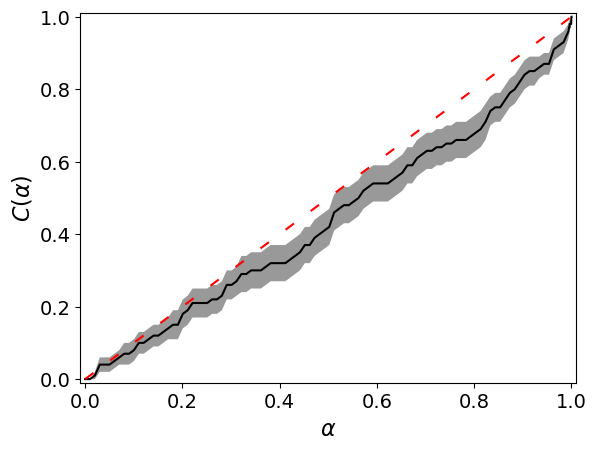}
    \includegraphics[width=\columnwidth]{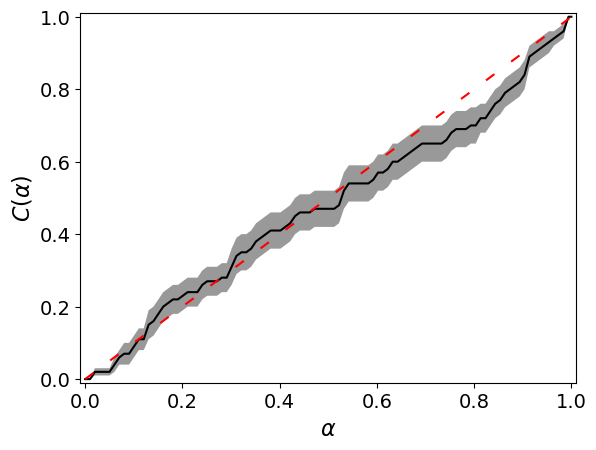}
    \caption{(Left) Coverage plot from the inference test from 100 models at $z = 5.4$ uniformly sampled from our priors on $T_0$, $\gamma$, and $\langle F \rangle$. (Right) Coverage plot derived from the inference test using 100 datasets generated from a Gaussian distribution with the mean model and covariance matrix at $z = 5.4$. }\label{fig: gaussian 5.4}
\end{figure*}

\begin{figure*}
   \includegraphics[width=\columnwidth]{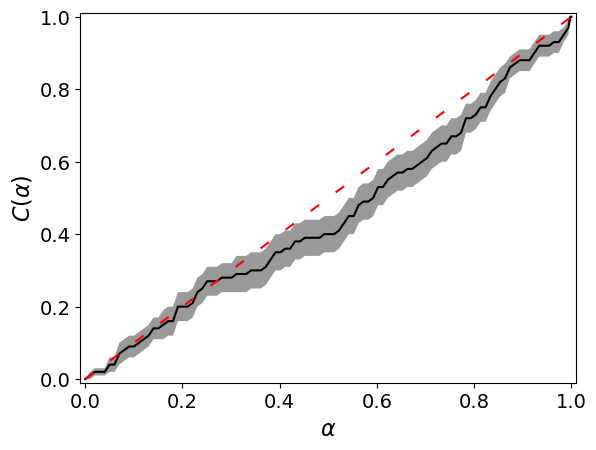}
    \includegraphics[width=\columnwidth]{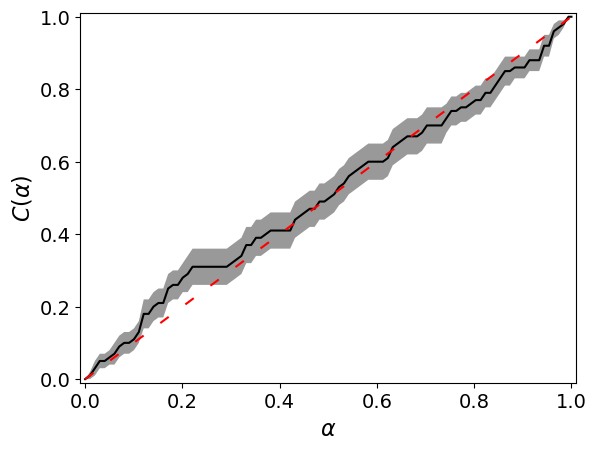}
    \caption{(Left) Coverage plot from the inference test from 100 models at $z = 5.7$ uniformly sampled from our priors on $T_0$, $\gamma$, and $\langle F \rangle$. (Right) Coverage plot derived from the inference test using 100 datasets generated from a Gaussian distribution with the mean model and covariance matrix at $z = 5.7$. }\label{fig: gaussian 5.7}
\end{figure*}

As discussed in Section~\ref{sec:inf_test}, we generate Gaussian-distributed data and perform the inference test to eliminate the impact of the non-Gaussian distribution present in the random mock data from forward-modelled sightlines. For one thermal model of $T_0, \gamma, \text{and} \langle F \rangle$, a random mock data set is sampled from the multi-variate Gaussian distribution of the mean model and model-dependent covariance matrix as described in Section~\ref{corr}, \ref{sec: covar}. Following the same inference test procedure as in Section~\ref{sec:inf_test}, we obtain results for $z = 5.4$ and $z = 5.7$ in Figure~\ref{fig: gaussian 5.4},~\ref{fig: gaussian 5.7}. Here both coverage plots fall along the $C(\alpha) = \alpha$ red dash line and remove the slight over-confident deviation (coverage plot goes under the red dash line) as expected. The same performance can be seen at other $z$. This validates that the non-Gaussian distribution of mock data of \lya~autocorrelation function at high $z$ leads to the offset in inference test. We therefore conclude that the inference procedure with the NN emulator is robust despite of the suboptimal choice of likelihood.

\subsection{Coverage Plots at Other Redshifts}
\label{appdix:cov_plots}
\begin{figure*} 
     \begin{subfigure}{0.45\textwidth}
        \includegraphics[width=\textwidth]{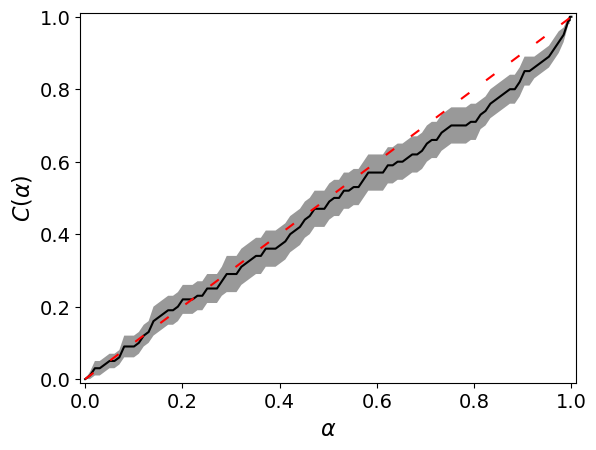}
         \caption{Coverage plot of 100 forward-modelled mocks at $z = 5.5$.}
     \end{subfigure}
      \hfill 
    \begin{subfigure}{0.45\textwidth}
        \includegraphics[width=\textwidth]{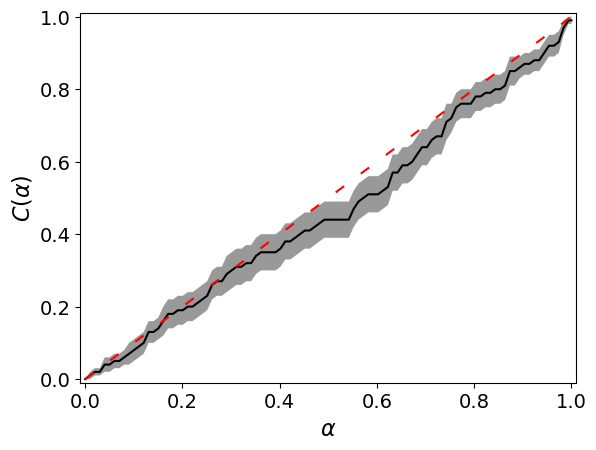}
         \caption{Coverage plot of 100 forward-modelled mocks at $z = 5.6$.}
     \end{subfigure}
     \hfill 
      \begin{subfigure}{0.45\textwidth}
        \includegraphics[width=\textwidth]{figures/coverage/z57_train_55_bin59_seed_11_inference_100_forward_mocks_emulator_seed_36_samples_4000_chains_4_nn_err_prop_True_test_12.png}
         \caption{Coverage plot of 100 forward-modelled mocks at $z = 5.7$.}
     \end{subfigure}
     \hfill 
    \begin{subfigure}{0.45\textwidth}
        \includegraphics[width=\textwidth]{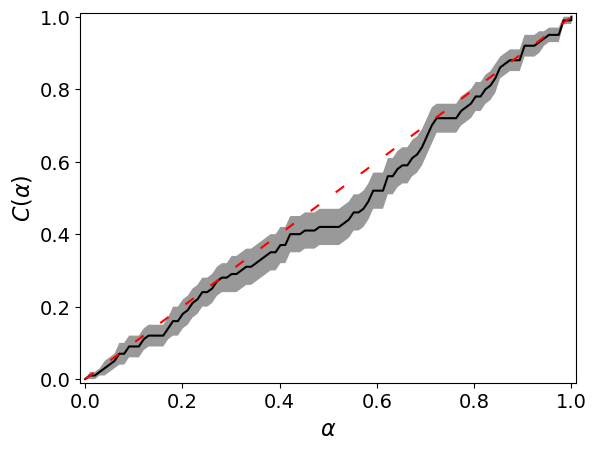}
         \caption{Coverage plot of 100 forward-modelled mocks at $z = 5.8$.}
     \end{subfigure}
     \hfill 
    \begin{subfigure}{0.45\textwidth}
        \includegraphics[width=\textwidth]{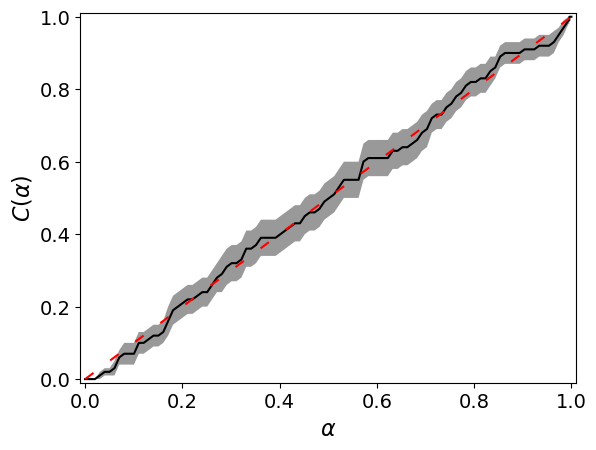}
         \caption{Coverage plot of 100 forward-modelled mocks at $z = 5.9$.}
     \end{subfigure}
     \hfill 
      \begin{subfigure}{0.45\textwidth}
        \includegraphics[width=\textwidth]{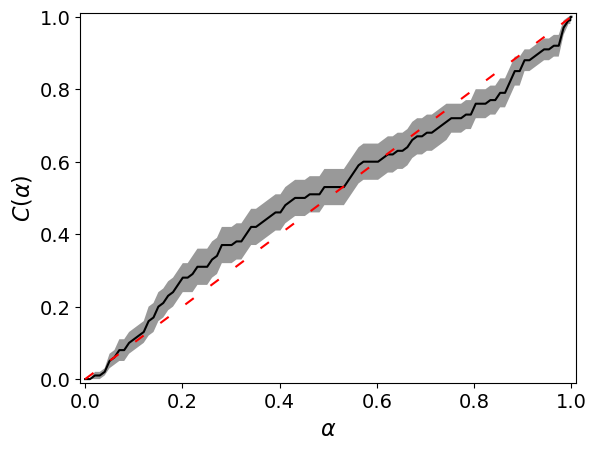}
         \caption{Coverage plot of 100 forward-modelled mocks at $z = 6.0$.}
     \end{subfigure}
\caption{This figure shows the coverage resulting from the inference test from 100 \lya~autocorrelation function mocks at $z = 5.5-6.0$ drawn from our priors on $T_0$, $\gamma$, and $\langle F \rangle$. Same thing for $z = 5.4$ as the orange contour in Figure~\ref{fig:compare_cov}. }\label{fig: all z coverage}
\end{figure*}

This section shows the inference test results for random forward-modelled mocks at other $z$ which can be compared to the orange contour in Figure~\ref{fig:compare_cov} for $z = 5.4$. Figure \ref{fig: all z coverage} shows that, at other $z$, the inference test results follow the same slight over-confident offset resulting from the non-Gaussianity of mock data. For higher $z$, because the $\langle F \rangle$ is lower, even though the distribution of sightline correlation values is more skewed \citepalias[see][appendix C]{wolfson2023forecasting}, the autocorrelation function shape is not affected much by the IGM temperature. The posteriors are therefore broader for higher redshift, $z = 5.9 - 6.0$, and our error-propagated inference method further broaden the posteriors leading to passing the inference test.

%%%%%%%%%%%%%%%%%%%%%%%%%%%%%%%%%%%%%%%%%%%%%%%%%%
% Don't change these lines
\bsp	% typesetting comment
\label{lastpage}
\end{document}